\documentclass{article}

\usepackage{microtype}
\usepackage{graphicx}
\usepackage{subfigure}
\usepackage{booktabs} 
\usepackage{makecell}
\usepackage{hyperref}

\usepackage[accepted]{sysml2019}

\sysmltitlerunning{Efficient and Robust Parallel DNN Training through Model Parallelism on Multi-GPU Platform}

\begin{document}

\twocolumn[
\sysmltitle{Efficient and Robust Parallel DNN Training through Model Parallelism on Multi-GPU Platform}




\begin{sysmlauthorlist}
\sysmlauthor{Chi-Chung Chen}{NTU}
\sysmlauthor{Chia-Lin Yang}{NTU}
\sysmlauthor{Hsiang-Yun Cheng}{AS}

\end{sysmlauthorlist}

\sysmlaffiliation{NTU}{National Taiwan University, Taipei, Taiwan}

\sysmlaffiliation{AS}{Academia Sinica, Taipei, Taiwan}

\sysmlcorrespondingauthor{Chia-Lin Yang}{yangc@csie.ntu.edu.tw}

\sysmlkeywords{Machine Learning, SysML}

\vskip 0.3in

\begin{abstract}
The training process of Deep Neural Network (DNN) is compute-intensive, often taking days to weeks to train a DNN model. 
Therefore, parallel execution of DNN training on GPUs is a widely adopted approach to speed up the process nowadays. 
Due to the implementation simplicity, data parallelism is currently the most commonly used parallelization method. 
Nonetheless, data parallelism suffers from excessive inter-GPU communication overhead due to frequent weight synchronization among GPUs. 
Another approach is pipelined model parallelism, which partitions a DNN model among GPUs, and processes multiple mini-batches concurrently. 
This approach can significantly reduce inter-GPU communication cost compared to data parallelism. 
However, pipelined model parallelism faces the weight staleness issue; that is, gradients are computed with stale weights, leading to training instability and accuracy loss.
In this paper, we present a pipelined model parallel execution method that enables high GPU utilization while maintaining robust training accuracy via a novel weight prediction technique, SpecTrain. Experimental results show that our proposal achieves up to 8.91x speedup compared to data parallelism on a 4-GPU platform while maintaining comparable model accuracy. 
\end{abstract}
]

\section{Introduction}

Deep Neural Networks (DNNs) have gained a lot interests in recent years as it has demonstrated success for many classification and regression tasks, including image recognition \cite{alexnet,vgg,residualnet,inceptionv4, DBLP:conf/aaai/RealAHL19}, language translation \cite{transformer,seq2seq,DBLP:journals/corr/BahdanauCB14, DBLP:journals/corr/abs-1907-05019} and speech recognition \cite{residuallstm,hinton2012deep}. 
Training a DNN model is compute-intensive and often takes days to weeks, due to the large amount of training data and the ever-increasing model size. 
As such, multi-GPU platforms are widely adopted to speed up DNN training through parallel execution. \cite{pipedream,DBLP:conf/nips/DeanCMCDLMRSTYN12,DBLP:conf/hotos/CiparHKLGGKX13,DBLP:conf/ijcai/ZhangGL016,DBLP:journals/corr/abs-1712-01887,DBLP:conf/nips/LianHLL15,DBLP:journals/corr/ChenMBJ16}. 

In multi-GPU platforms, data parallelism \cite{DBLP:journals/corr/Krizhevsky14} is a commonly used parallelization approach.
When data parallelism is applied, each GPU holds a complete copy of DNN model and processes a subset of the training data.
Data parallelism suffers from excessive inter-GPU communication overheads, since weights updated at each individual GPU need to be synchronized. 
These communication overheads increase as the model size increases, severely hindering the scalability of data parallelism \cite{DBLP:conf/interspeech/SeideFDLY14,DBLP:conf/interspeech/Strom15,DBLP:journals/corr/Alistarh0TV16,DBLP:journals/corr/ZhouNZWWZ16,DBLP:conf/emnlp/AjiH17,DBLP:journals/corr/abs-1712-01887}.

   
Another parallelization approach is model parallelism, which partitions a DNN model among GPUs. Each GPU is responsible for the weight updates of the assigned model layers.  Therefore, the amount of data communicated among GPUs is significantly less than data parallelism.  Furthermore, model parallelism enables training a large model exceeding the size constraint limited by the accelerator memory capacity. However, for a naive model parallelism implementation, only one GPU is active at a time. To enable parallel execution, PipeDream \cite{pipedream} proposes to adopt pipelining by injecting multiple mini-batches to the model concurrently.  However, pipelined model parallelism introduces the staleness and consistency issue for weight updates.  Since multiple mini-batches are simultaneously processed in the pipeline, a later mini-batch could start the training process before its prior mini-batch updates weights. The staleness/consistency problem leads to unstable training and degrades model accuracy. Pipedream\cite{pipedream} seeks to alleviate the consistency issue by ensuring the forward and backward passes of a mini-batch in the same GPU sees the same weights through storing multiple versions of weights. However, the weight staleness problem still remains unsolved. To completely avoid the weight staleness and consistency issue, Gpipe \cite{DBLP:journals/corr/abs-1811-06965} proposes a different pipelining method from Pipedream. It divides each mini-batch into multiple micro-batches and pipelining the micro-batches of the same mini-batch, and weights are updated synchronously at the end of a mini-batch. Even though GPipe could increase GPU utilization compared with non-pipelining model parallelism, it achieves significantly lower GPU throughput than Pipedream. According to the experiment in \cite{pipedream_thesis}, the throughput of GPipe could be only 29\% of Pipedream.

In this paper, we present a pipelined model parallel execution method that enables high GPU utilization while maintaining robust training accuracy via a novel weight prediction technique, SpecTrain. 
We adopt the pipelining method in Pipedream and perform detailed comparison with data-level parallelisms on a multi-GPU platform.  For CNN models, pipelined model parallelism performs comparable to data parallelism. But for FCN/RNN models which have massive model parameters, pipelined model parallelism shows superior performance advantage over data parallelism , up to 8.91x speedup on a 4-GPU platform. However, in the aspect of training robustness, pipelined model parallelism shows unstable learning and leads to worse model accuracy than data parallelism due to the weight staleness problem.  To tackle this challenge, the proposed SpecTrain mechanism, predicts the future weights in early pipeline stages, so that a mini-batch can adopt the predicted future weights rather than the stale weights to perform its computation. 
The design is based on the observation that smoothed gradients used in momentum-based optimizers \cite{adam} reflect the trend of weight updates and can be leveraged for accurate weight prediction. Our evaluations show that SpecTrain induces no accuracy drop in most of the workloads compared to data parallelism, while Pipedream incurs 1.1\% accruacy drop. 


The rest of the paper is organized as follows.
Section~\ref{sec:background} provides background on parallel DNN training and compare data parallelism and model parallelism in terms of load balance, inter-GPU communications, and training efficiency.
The staleness issue and our weight prediction technique, SpecTrain, are introduces in Section~\ref{sec:dualpipe}.
Section~\ref{sec:exp} describes evaluation methodologies and results, followed by a summary of related work in Section~\ref{sec:relatedwork}.
Finally, we conclude the paper in Section~\ref{sec:conclusion}.

\section{Background and Motivation}
\label{sec:background}

\begin{figure}
    \centering
    \includegraphics[width=0.35\textwidth]{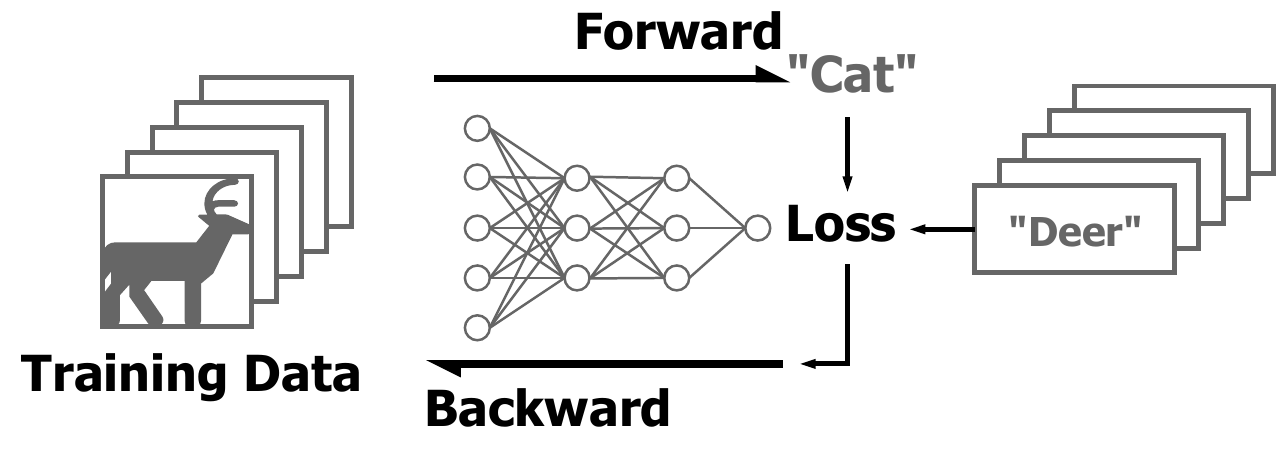}
    \caption{Overview of Stochastic Gradient Descent.}
    \label{fig:train_overview}
\end{figure}

\begin{figure}
    \centering
    \subfigure[Data Parallelism.] {
        \centering
        \includegraphics[width=0.35\textwidth]{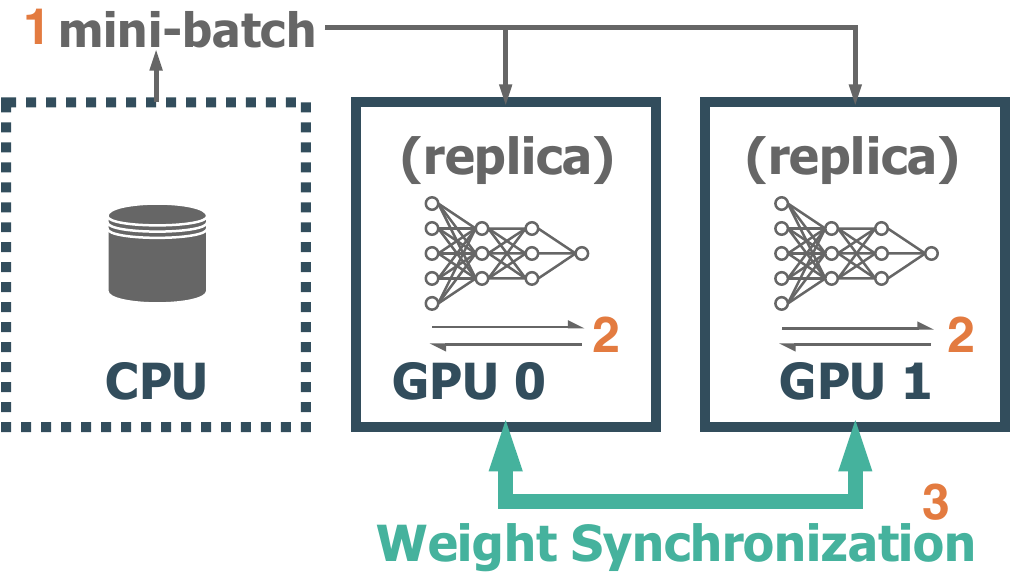}
        \label{fig:dp_overview}
    }
    \newline
    \subfigure[Model Parallelism.] {
        \centering
        \includegraphics[width=0.35\textwidth]{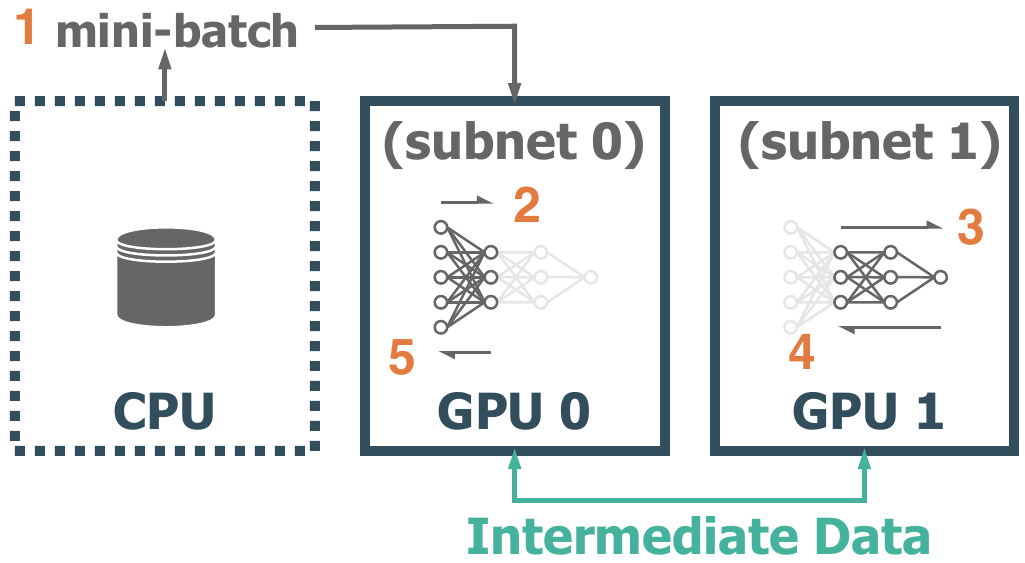}
        \label{fig:mp_overview}
    }
    \caption{Training process with different parallelization methods.}
\end{figure}

\subsection{DNN Training}

DNN models are typically trained using Stochastic Gradient Descent (SGD), as shown in Figure~\ref{fig:train_overview}. 
Training data are randomly sampled into mini-batches. 
Mini-batches are fed into the model to traverse the model in two phases: forward and backward passes.
The forward pass generates predictions and calculates the loss between the prediction and the ground truth. 
Then the backward pass backpropagates errors to obtain gradients to update model weights. 
A mini-batch passing though the forward and backward phases is referred to as an iteration, and an epoch is defined as performing the forward-backward pass through the entire training dataset. 
The training process iterates multiple epochs until the model converges. 

\subsection{Parallel Training} \label{subsec:parallel}

The training process of DNN is time-consuming, taking days or weeks to finish a large-scale training project. 
Parallelizing DNN training is needed to speedup the process. 
There are two parallelization approaches: data parallelism and model parallelism. 

For data parallelism \cite{DBLP:journals/corr/Krizhevsky14}, each GPU holds a full copy of the model and processes a different set of training data, as shown in Figure~\ref{fig:dp_overview}. 
Each GPU computes its own gradients. 
These gradients are aggregated at the parameter server by summation. 
The aggregated gradients are then broadcasted to all GPUs to update weights. 
Since every GPU receives the same gradients, weights in the DNN model stay consistent among GPUs. 
The gathering and scattering process of gradients is called weight synchronization, which requires a large amount of inter-GPU communications. 
A variant of data parallelism widely used in distributed systems is asynchronous data parallelism \cite{DBLP:conf/nips/DeanCMCDLMRSTYN12,DBLP:conf/hotos/CiparHKLGGKX13,DBLP:conf/ijcai/ZhangGL016,DBLP:conf/nips/LianHLL15,DBLP:journals/corr/ChenMBJ16}. 
When asynchronous data parallelism is applied, the parameter server is in charge of weight updates. 
Each GPU sends its gradients to the parameter server, which then updates weights and send the updated weights back to that GPU.
In this way, there is no synchronization among GPUs. 
This approach solves the unstable networking issue in distributed computing environment but it introduces the inconsistency issue. 
Moreover, this method does not reduce the amount of data transfers among GPUs. 
Since we target on multi-GPU systems where GPUs are connected with PCIe links, synchronous data parallelism is preferred. 

Model parallelism \cite{pipedream} partitions a model among multiple GPUs, where each GPU is responsible for the weight updates of the assigned model layers, as shown in Figure~\ref{fig:mp_overview}. 
Intermediate data like layer outputs for the forward pass and gradients for the backward pass are transferred among GPUs. 
Since these partitions have dependencies, in a naive implementation of model parallelism, only one GPU is active at a time, leading to low GPU utilization. 
To enable parallel execution, PipeDream \cite{pipedream} proposes to adopt pipelining by injecting multiple mini-batches to the pipeline concurrently. 
Therefore, each GPU could process different mini-batches simultaneously. 

\subsection{Data Parallelism vs. Model Parallelism} \label{subsec:dp_vs_mp}

We compare data parallelism and model parallelism in three aspects: load balance across GPUs, inter-GPU communication, and training efficiency. 

\paragraph{Load Balance across GPUs}

Data parallelism partitions training data across multiple GPUs, therefore, load balance could be easily maintained. 
As for model parallelism, achieving load balance is more challenging. 
Since the complexity of different DNN layers varies, it would introduce significant efforts for programmers to partition model layers to GPUs in a balanced way. 
A few prior works have addressed this issue\cite{pipedream,DBLP:conf/icml/MirhoseiniPLSLZ17}. PipeDream \cite{pipedream} proposes to profile the processing time of each layer offline and use dynamic programming to partition the model. 
Mirhoseini et al. \cite{DBLP:conf/icml/MirhoseiniPLSLZ17} adopts reinforcement learning to dynamically partition the model at run-time. 

\paragraph{Inter-GPU Communication}

\begin{figure}[t]
    \centering
    \includegraphics[width=0.4\textwidth]{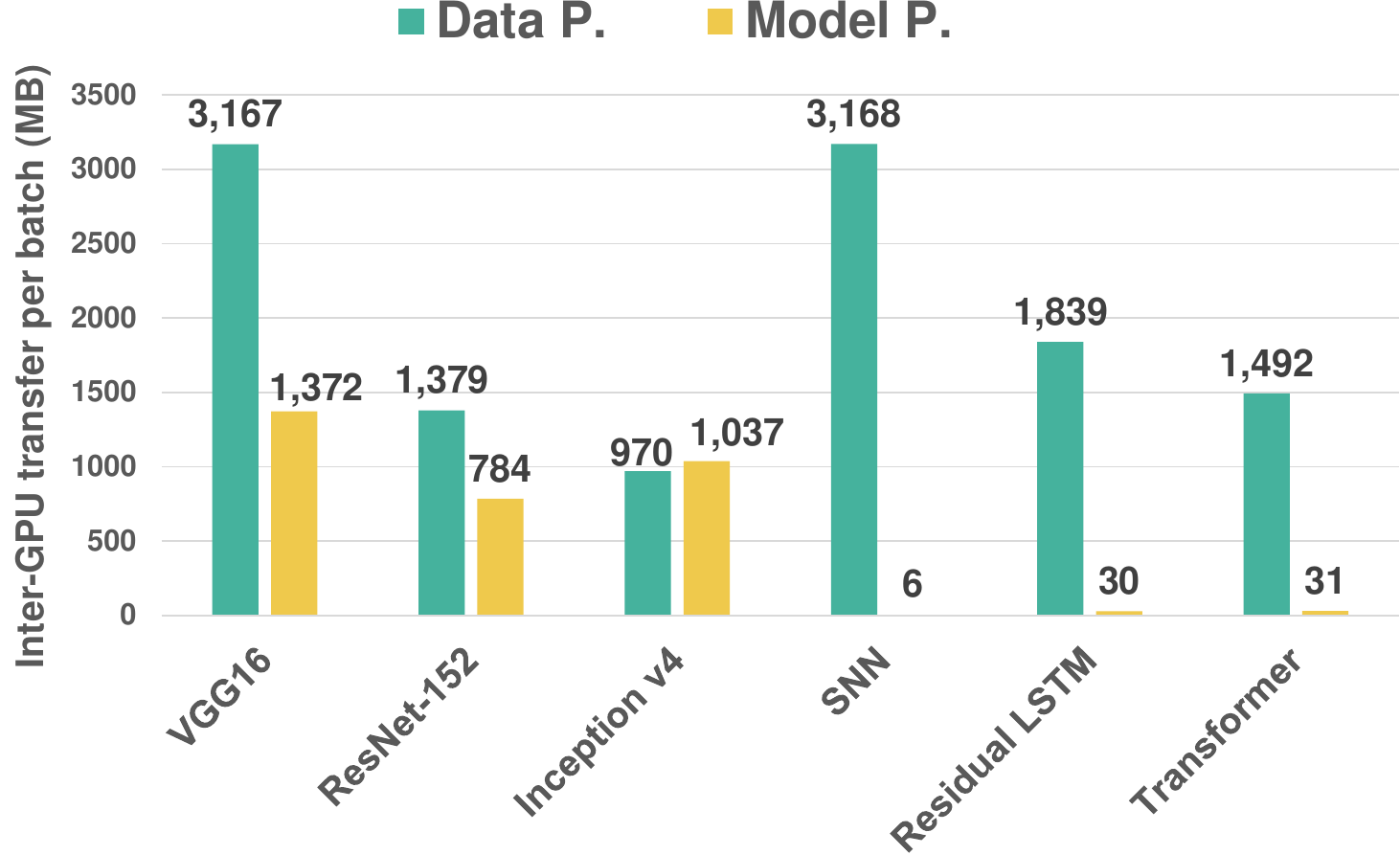}
    \caption{The amount of inter-GPU data transfers per mini-batch when applying different parallization approaches on a 4-GPU system.}
    \label{fig:p2p_transfer_amount}
\end{figure}

\begin{figure}[t]
    \centering
    \includegraphics[width=0.4\textwidth]{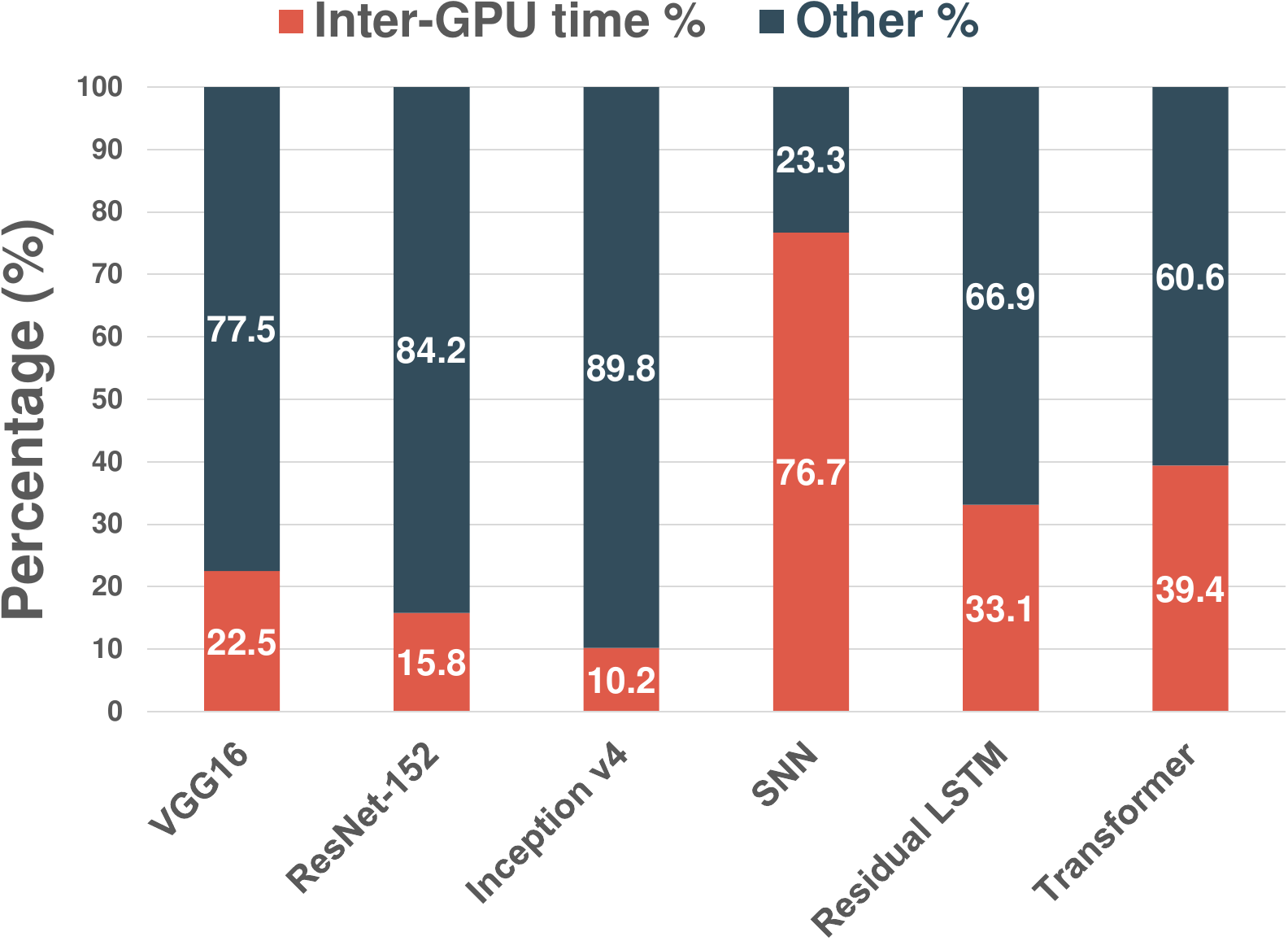}
    \caption{Percentage of time spent on inter-GPU communications when applying data parallelism on a 4-GPU system.}
    \label{fig:dp_breakdown}
\end{figure}

Both data parallelism and model parallelism need inter-GPU communications. 
Data parallelism communicates gradients for weight synchronization, while model parallelism transfers intermediate data between sub-models. 
Figure~\ref{fig:p2p_transfer_amount} compares the amount of data transfers among GPUs for these two parallization methods.\footnote{The experimental setup is described in Section~\ref{subsec:exp_setup}.} 
We observe that data parallelism requires 13.4x more inter-GPU communications on average (up to 528x) than model parallelism. 
The only exception is \emph{Inception v4} \cite{inceptionv4}, as its relatively smaller size of weights reduces the cost of weight synchronization.
\emph{SNN} \cite{snn}, \emph{Residual LSTM} \cite{residuallstm}, and \emph{Transformer} \cite{transformer} require almost no communications between GPUs when model parallelism is applied, since these models have fewer intermediate data between layers. 
The excessive inter-GPU communications of data parallelism lead to considerable slowdown. 
As shown in Figure~\ref{fig:dp_breakdown}, on average, 26.7\% of training time is spent on inter-GPU data transfer (up to 76.7\%) when data parallelism is applied. 

\paragraph{Training Efficiency}

\begin{figure}
    \centering
    \includegraphics[width=0.4\textwidth]{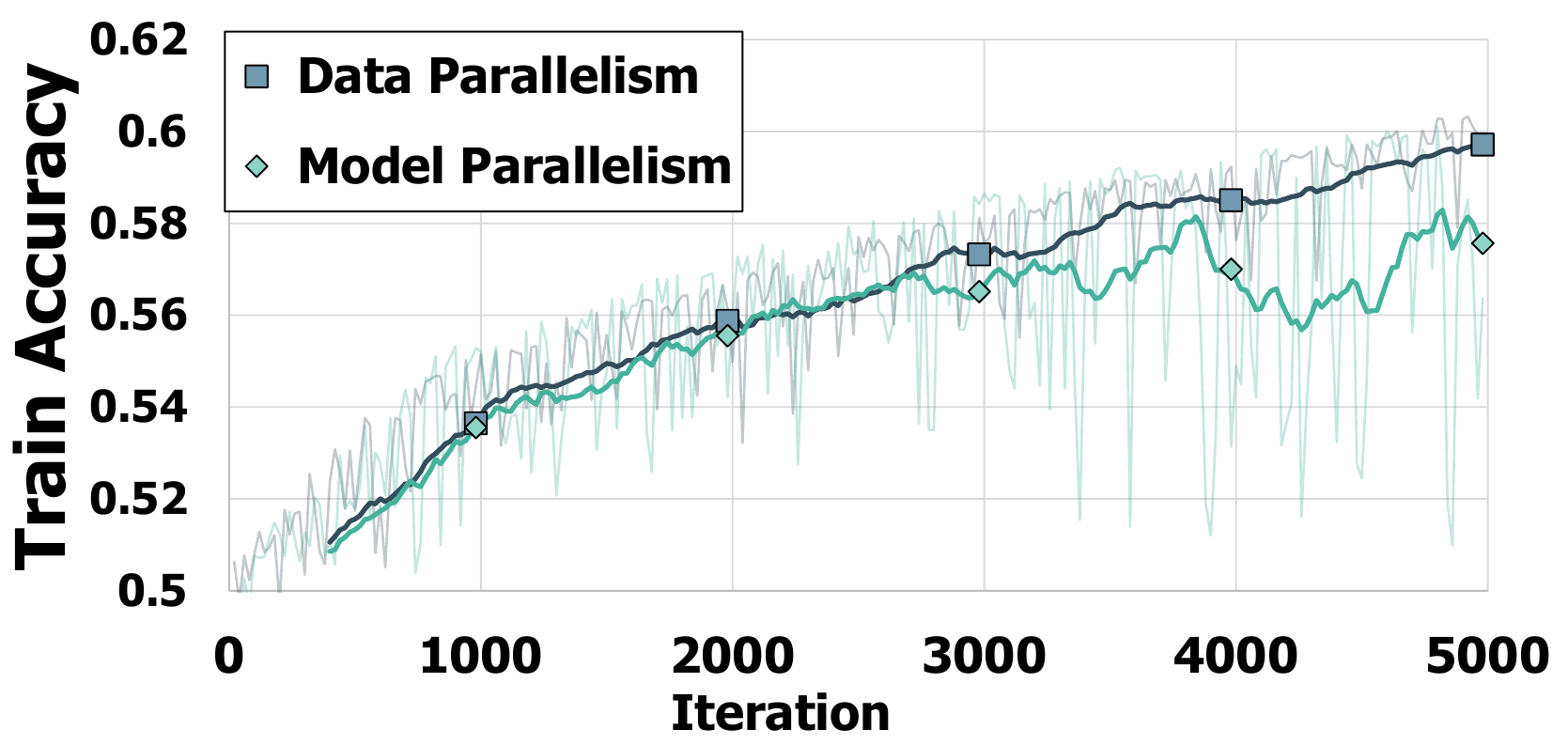}
    \caption{The accuracy curve when training \emph{Transformer} \cite{transformer}. The light-color curve denotes the measured values. To make the trends clear, we add a thick line to show the moving average over 20 latest values.}
    \label{fig:transformer_pre_accuracy}
\end{figure}

Both data parallelism and model parallelism affect DNN training efficiency, i.e., the model convergence rate and model accuracy.
For data parallelism, data partition size affects a key DNN training parameter: the mini-batch size.
A large mini-batch size increases the GPU utilization but could degrade the model accuracy if it is excessively large \cite{DBLP:journals/corr/WuZE17, DBLP:journals/corr/KeskarMNST16}. In contrast, a small mini-batch size may under-utilize GPU resources. 
For data parallelism, if each GPU performs the training process for a mini-batch, then the effective mini-batch size is the number of GPU times the mini-batch size. 
To avoid the adverse effect of a large mini-batch size, we could choose to partition a mini-batch among GPUs. 
However, it could under-utilize GPU resources. 
As the number of GPU increases, it becomes more and more challenging to select a data partition size that is good for DNN training efficiency.

For model parallelism, if training is proceeded in the pipelined manner, it induces the staleness issue. 
Since multiple mini-batches are in progress in the pipeline, before earlier mini-batches update weights, latter mini-batches adopt stale weights to derive gradients. 
This staleness issue leads to unstable learning and worse model accuracy. Figure~\ref{fig:transformer_pre_accuracy} shows the comparison of model accuracy, i.e., the percentage of correct classifications, between model and data parallelism. 
We observe that the accuracy of data parallelism steadily increases as training goes on, but the accuracy of model parallelism fluctuates.
 
As the DNN model continues to grow, we need to deploy more GPUs to speedup training.
As such, data parallelism is expected to face the scalability issue.
Model parallelism appears to be appealing for parallelizing DNN training. 
Therefore, in this paper, we tackle the main challenge, i.e., the staleness issue, for realizing a robust and efficient model parallelism approach to parallelize DNN training.

\section{Robust Pipeline Design} \label{sec:dualpipe}


\subsection{Staleness Issue}

\begin{figure}
    \centering
    \includegraphics[width=0.4\textwidth]{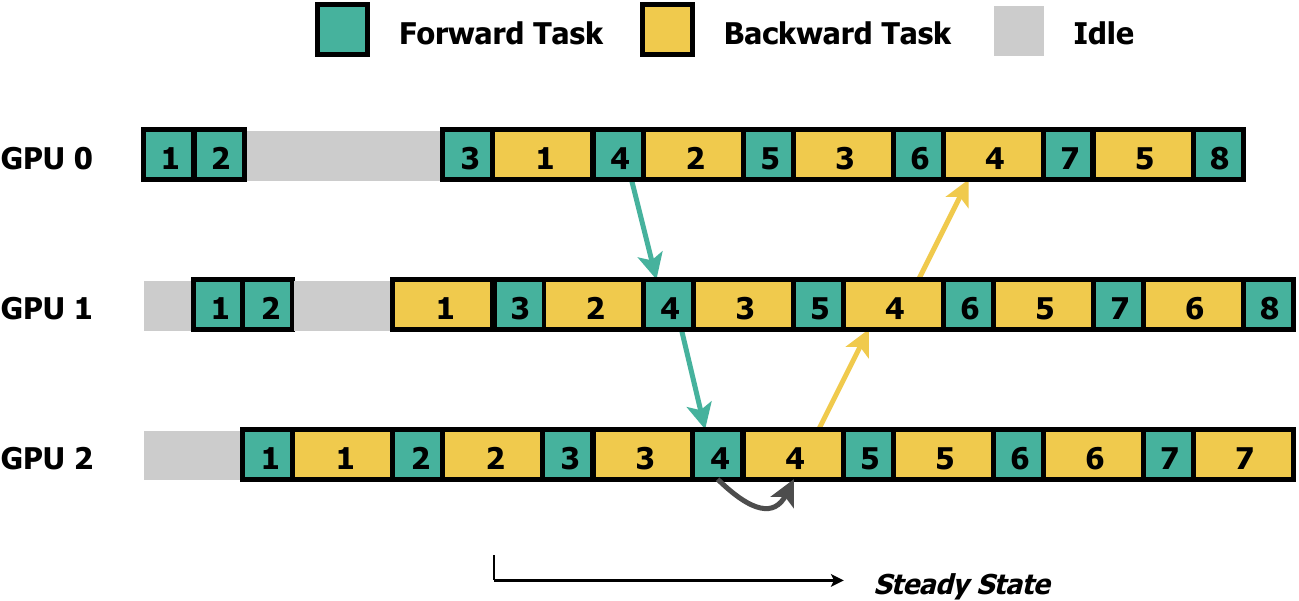}
    \caption{Example timeline of pipelined model parallelism. The boxes with index $i$ denotes the time spent on processing the $i$-th mini-batch. Cyan and yellow boxes indicate forward and backward tasks respectively. A round trip of processing a mini-batch is presented by the arrows.}
    \label{fig:timeline}
\end{figure}

Pipelined DNN training is bi-directional \cite{pipedream}; that is, a mini-batch flows through the pipeline (forward pass) and then traverse back for weight updates (backward pass).
It issues a forward task and a backward task in a round-robin manner as demonstrated in Figure~\ref{fig:timeline}. 
Once a task is completed, it asynchronously executes the next one to maximize the utilization.
The output data will be sent to the next GPU via a background thread to hide the latency.

\begin{figure}[t]
    \centering
    \subfigure[Single GPU]{
        \includegraphics[width=0.4\textwidth]{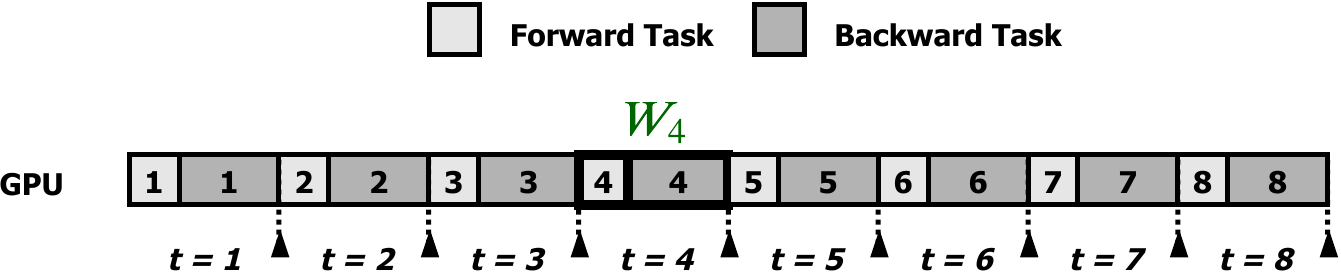}
        \label{fig:weightver_single}
    }
    \newline
    \subfigure[Pipelining]{
        \includegraphics[width=0.4\textwidth]{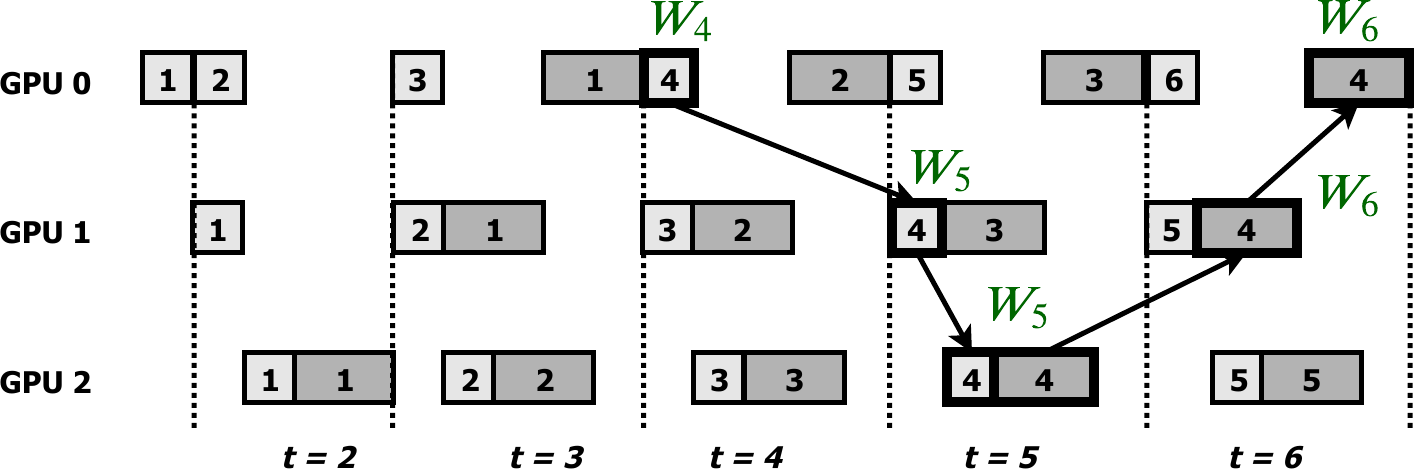}
        \label{fig:weightver_pipeline}
    }
    \newline
    \subfigure[Pipelining with PipeDream Weight Stashing.]{
        \includegraphics[width=0.4\textwidth]{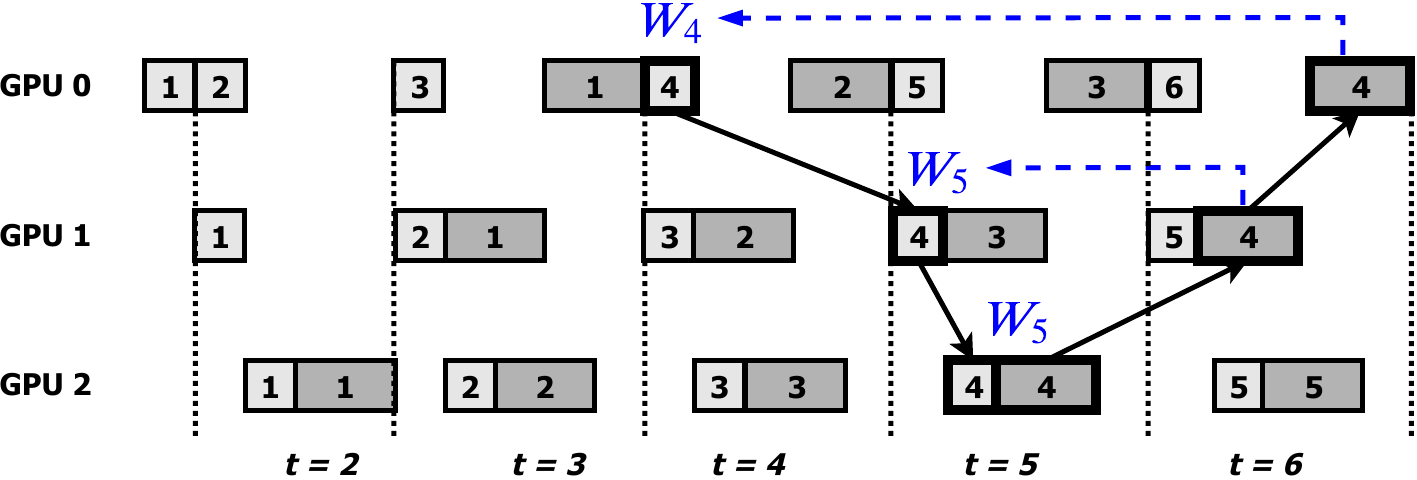}
        \label{fig:weightver_pipedream}
    }
    \newline
    \subfigure[Pipelining with SpecTrain]{
        \includegraphics[width=0.4\textwidth]{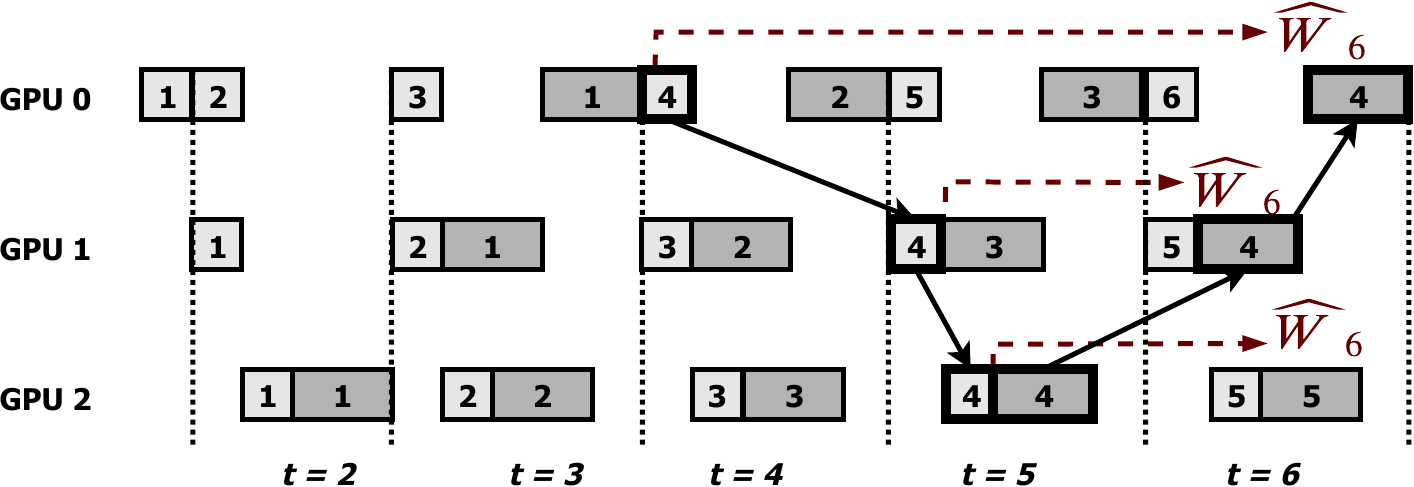}
        \label{fig:weightver_spectrain},
    }
    \caption{Example timeline of single-GPU, pipelining, PipeDream with weight stashing, and our SpecTrain. The round trip of the 5-$th$ mini-batch and its adopted weight versions are illustrated. Note that we intentionally align tasks among GPUs to mark the timestamps. In actual execution, GPUs run the next task back to back without waiting other GPUs.}
    \label{fig:weightver}
    
\end{figure}

Staleness is a critical issue that should be resolved in pipelined model parallelism.
We use an example, as shown in Figure~\ref{fig:weightver}, to explain the staleness issue in pipelining. 
In this example, $W_{t}$ represents the version of weights at time unit $t$.

In the staleness-free single GPU implementation of training, both the forward and backward passes of a mini-batch are performed based on the latest weights.
At time $t$, the forward and backward passes of a mini-batch both perform computation based on the latest weights $W_{t}$, which was generated by the backward pass of the previous mini-batch at time $t-1$.
For example, in Figure~\ref{fig:weightver_single}, the 3-$rd$ mini-batch produces $W_{4}$, and the processing of the 4-$th$ mini-batch at the 4-$th$ time unit is based on $W_{4}$.

Different from the single GPU implementation, mini-batches in pipelined training adopt inconsistent and stale weights to perform computation. 
In pipelined training, a mini-batch is processed by different GPUs in multiple consecutive time units to finish the forward and backward passes.
Since multiple mini-batches are in progress in the pipeline, weights are continuously updated at every time unit.
Thus, a mini-batch adopts inconsistent versions of weights during its round trip (i.e., the entire flow of forward and backward passes) in the pipeline.
In addition, before earlier mini-batches update the weights, a latter mini-batch performs computation based on stale weights.
For example, in Figure~\ref{fig:weightver_pipeline}, the 4-$th$ mini-batch adopts various versions of weights, ranging from $W_{4}$ to $W_{6}$, during its round trip.
From the 4-$th$ mini-batch's perspective, $W_{4}$ and $W_{5}$ are stale and the only staleness-free version of weights is $W_{6}$, as $W_{6}$ is derived after the 3-$rd$ mini-batch updates the weights.  Such a weight update behavior is called the weight staleness issue and it leads to unstable and inferior convergence.

\subsection{SpecTrain: Staleness Mitigation via Weight Prediction}

To resolve the staleness issue, we propose a weight prediction method, SpecTrain, and apply this method in the pipeline.
SpecTrain predicts future weights in early pipeline stages, so that a mini-batch can perform computation based on the predicted future weights rather than the stale weights.
Our goal is to build a consistent and staleness-free training procedure.

In a consistent training procedure, the entire round trip of a mini-batch should adopt the same weight version.
PipeDream~\cite{pipedream} proposes Weight Stashing to alleviate the inconsistency problem.
Weight Stashing ensures the forward and backward passes of a mini-batch in the same GPU uses the same weights as demonstrated in Figure~\ref{fig:weightver_pipedream}, where the 4-$th$ mini-batch uses $W_{4}$ during both forward and backward pass on GPU 0, even though newer versions of weights are produced. Pipedream applies Weight Stashing in their experiments.
To implement Weight Stashing, every GPU should maintain a queue for storing several old versions of weights.
This additional queue wastes GPU memory space. Moreover, the training process of Weight Stashing still uses stale weights. 

To maintain weight consistency and avoid staleness, SpecTrain predicts future weights and adopts the predicted weights, rather than the earliest version of weights, throughout the entire round trip of a mini-batch.
Figure~\ref{fig:weightver_spectrain} illustrates the idea of SpecTrain.
Suppose that a mini-batch completes its round trip at time $t$, at early pipeline stages, the mini-batch predicts the future version of weights ($\widehat{W}_{t}$), which is expected to become the most updated weights at time $t$, and uses this future version of weights to perform computation.
For example, in Figure~\ref{fig:weightver_spectrain}, the processing of the 4-$th$ mini-batch in its entire round trip is based on $\widehat{W}_{6}$ rather than $W_{4}$. 

\paragraph{Weight Prediction.}

\begin{figure*}
    \centering
    \subfigure[Version difference $s = 1$.] {
        \centering
        \includegraphics[width=0.3\textwidth]{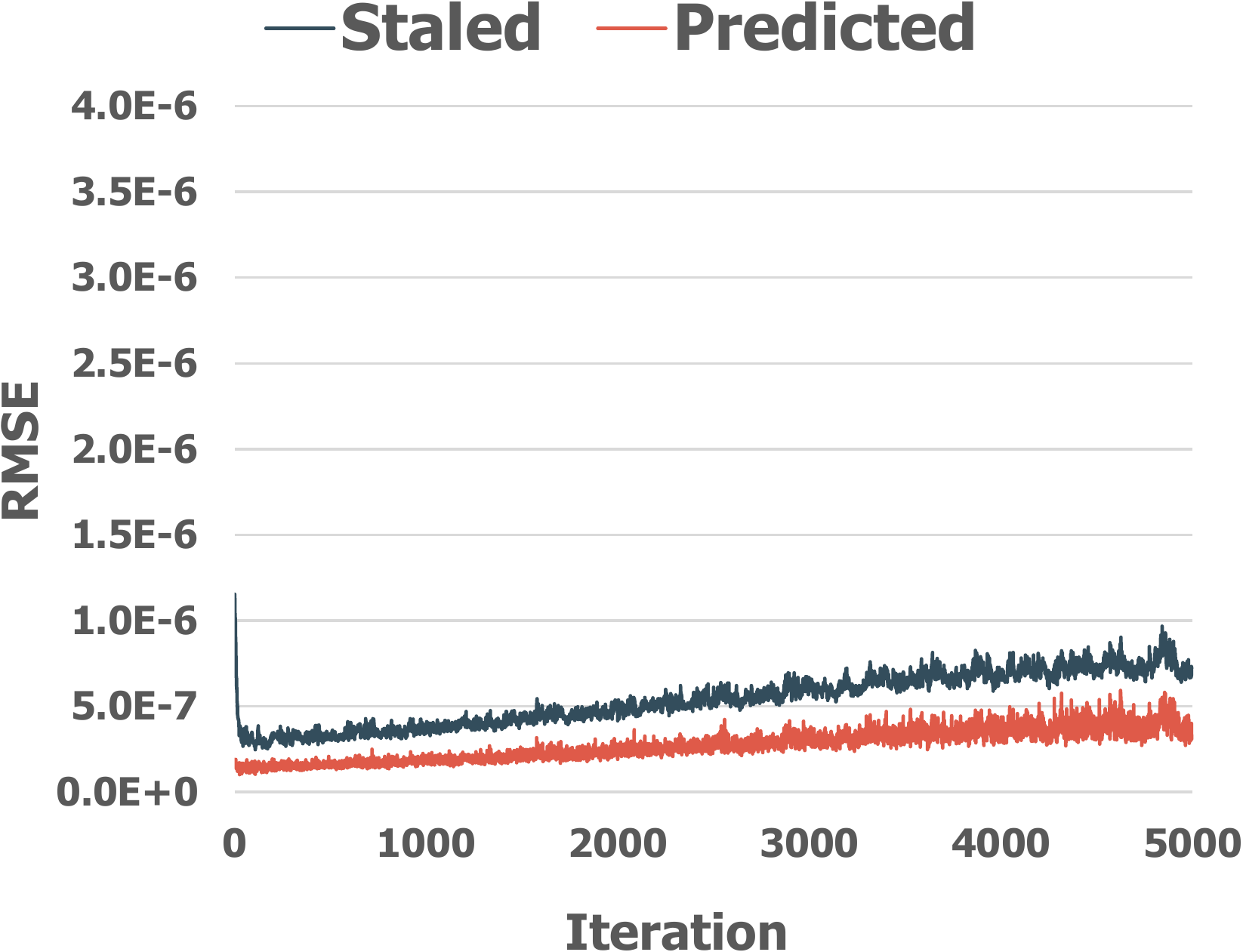}
    }
    \subfigure[Version difference $s = 2$.] {
        \centering
        \includegraphics[width=0.3\textwidth]{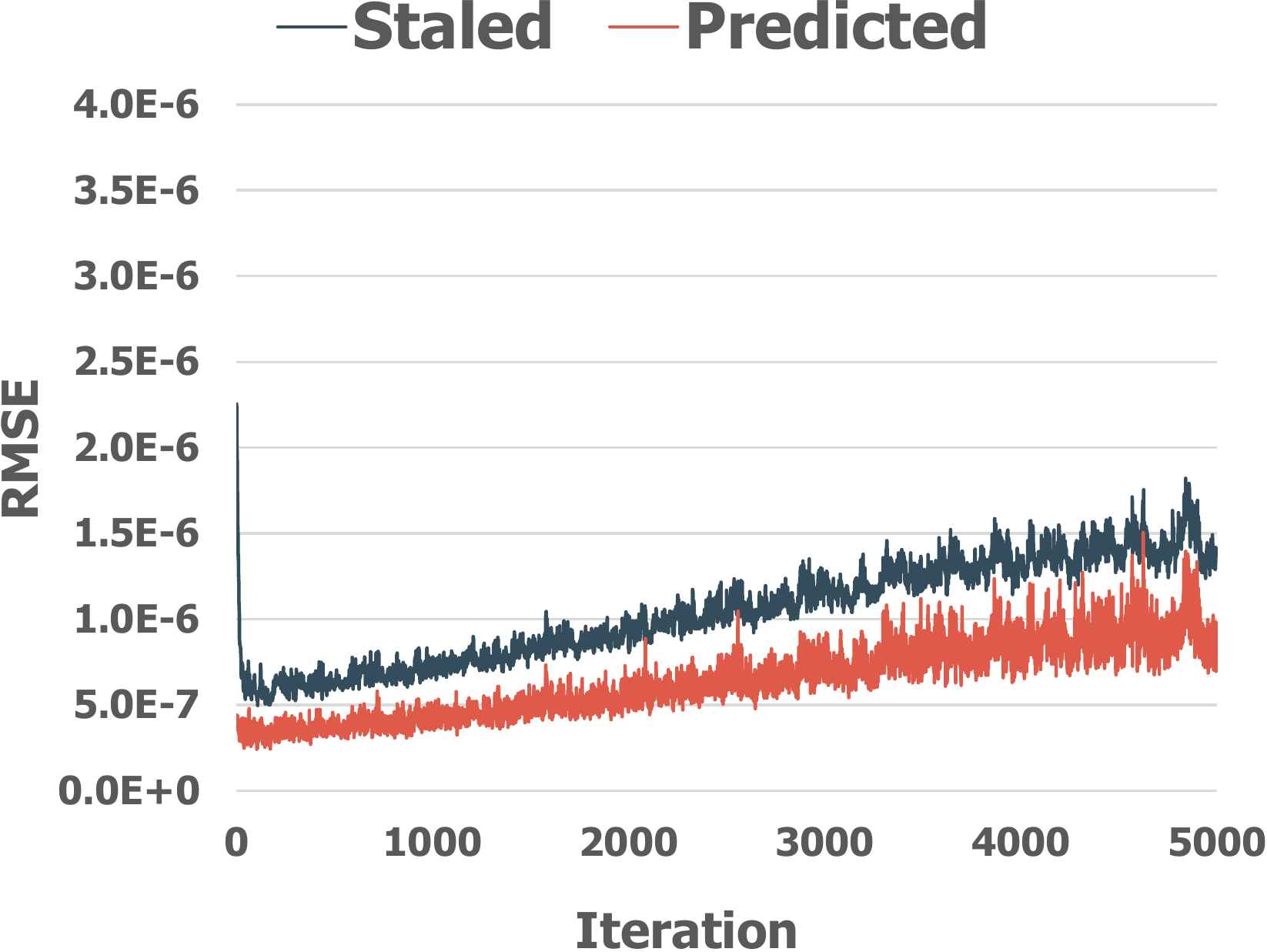}
    }
    \subfigure[Version difference $s = 3$.] {
        \centering
        \includegraphics[width=0.3\textwidth]{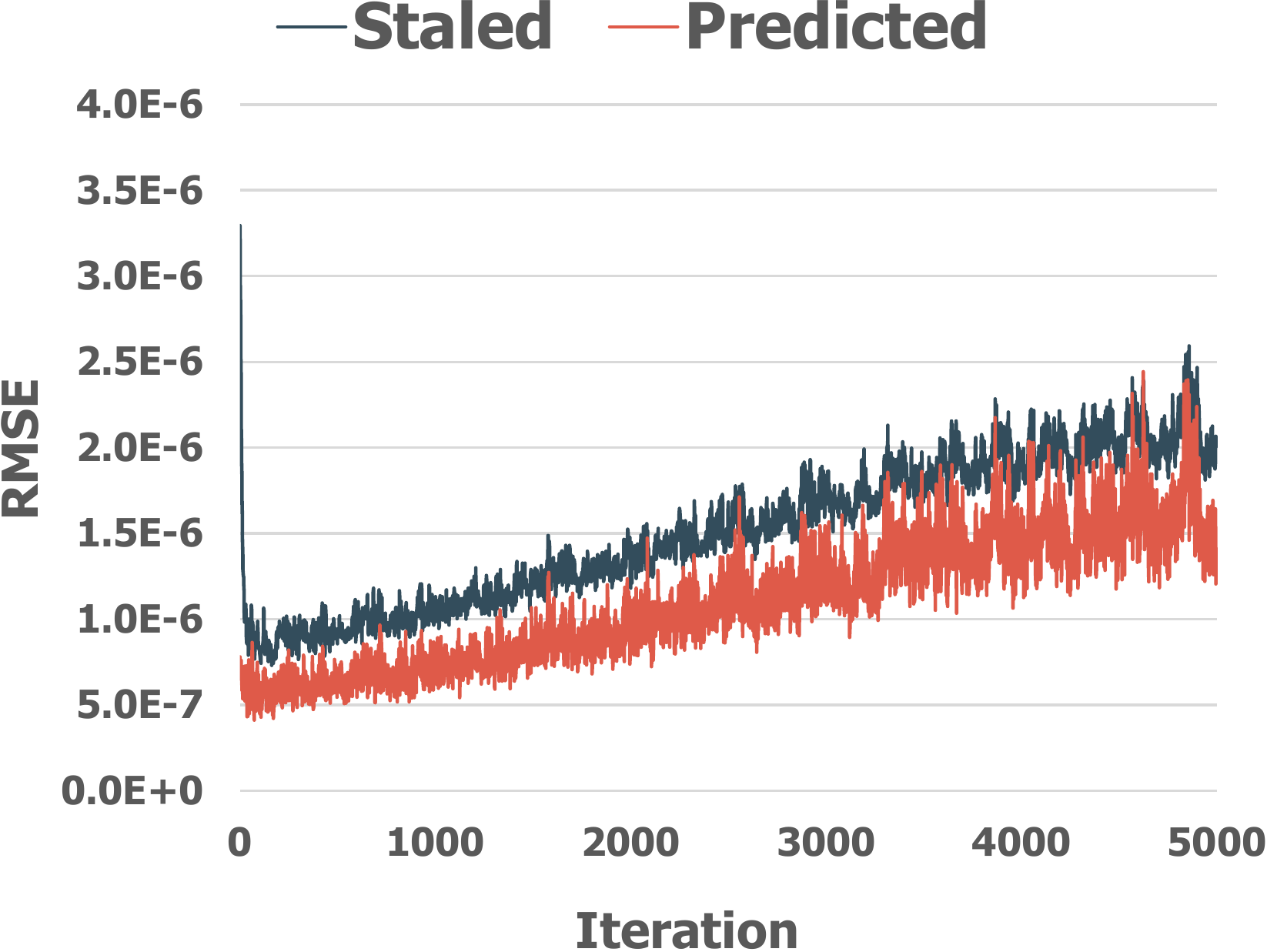}
    }
    \caption{Comparison between the RMSE of the predicted weights and the RMSE of the stale weights when training SNN on a 4-GPU system.}
    \label{fig:rmse}
\end{figure*}

SpecTrain predicts future weights based on the observation that smoothed gradients used in Momentum SGD~\cite{adam} reflect the trend of weight updates.
Momentum SGD is a common technique that can help to speed up and improve the stability of SGD by smoothing the weight updates.
A \textit{smoothed gradient} ($v_{t}$) is the weighted average of recent gradients and is calculated by the following equation:

\begin{equation}
    v_{t} = \gamma \cdot v_{t-1} + (1-\gamma) \cdot g_{t}
\end{equation}
where $\gamma$ is the decay factor with $0 < \gamma \leq 1$ and $g_{t}$ is the newly generated gradient.
Through averaging with recent gradients by the decay factor, smoothed gradient $v_{t}$ reflects the trend of weight updates.
Thus, we can use smoothed gradients to predict future weights.

In original SGD, weight update from time $t$ to time $t+1$ is conducted by the following equation:
\begin{equation} \label{eq:update}
    W_{t+1} = W_{t} - \eta \cdot g_{t}
\end{equation}
where $\eta$ is the learning rate.
Since our goal is to predict future weights, the true gradient $g_{t}$ is not derived yet when we make the prediction at an early pipeline stage.
Nevertheless, we can use smoothed gradient $v_{t-1}$ to replace the true gradient $g_{t}$, as smoothed gradient reflects the trend of weight updates.
Thus, we can use the following equation to make weight prediction:
\begin{equation} \label{eq:predict_single}
    \widehat{W}_{t+1} = W_{t} - \eta \cdot v_{t-1}
\end{equation}

Given a version of weights and the version difference $s$, we can make prediction on the future version of weights by recursively applying Equation~\ref{eq:predict_single}, as shown below:
\begin{equation} \label{eq:predict_s}
    \widehat{W}_{t+s} = W_{t} - s \cdot \eta \cdot v_{t-1}
\end{equation}

The version difference $s$ is defined as the number of time units between current pipeline stage and the time a mini-batch completes its round trip.
As shown in the following equations, the calculation of version difference $s$ depends on GPU index, $k$, and whether the mini-batch is at a forward or backward pass.
For mini-batches at forward passes, assuming that there are $N$ GPUs in the multi-GPU system, the version difference is calculated by

\begin{equation}\label{eq:s_forward}
    s = \lfloor k / 2 \rfloor + N - k - 1
\end{equation}

For mini-batches at backward passes, the version difference can be simply calculated by
\begin{equation}\label{eq:s_backward}
    s = \lfloor k / 2 \rfloor
\end{equation}
For example, in Figure~\ref{fig:weightver_spectrain}, at the 4-$th$ time unit, the 4-$th$ mini-batch is at its first pipeline stage of a forward pass and its version difference is $s = \lfloor 0 / 2 \rfloor + 3 - 0 - 1 = 2$, as this mini-batch will not complete until the 6-$th$ time unit.
Thus, the future weights of the 4-$th$ mini-batch can be predicted by $\widehat{W}_{6} = W_{4} - (6-4) \cdot \eta \cdot v_{3}$.
By using the predicted weights $\widehat{W}_{6}$ rather than the stale weights $W_{4}$ to perform computation, SpecTrain can mitigate the staleness issue in pipelined model parallelism.

\paragraph{Prediction Accuracy.}

We conducted an experiment to evaluate the accuracy of our weight prediction.
To quantify the accuracy, we calculate the root-mean-square error (RMSE)  between the predicted weights, $\widehat{W}_t$, and the actual weights, $W_t$.
As a comparison, the RMSE between the stale weights, $W_{t-s}$, and actual weights, $W_t$, is also analyzed.
The experiment is conducted by training the \emph{SNN}~\cite{snn} model~\footnote{Experimental setup is described in Section~\ref{subsec:exp_setup}. Due to space limitation, we only show the RMSE results of \emph{SNN}. Experiments on other models also show similar results.}.
To demonstrate how the version difference $s$ affects the prediction accuracy, we perform RMSE evaluations on $s=1$, $s=2$ and $s=3$.
Figure~\ref{fig:rmse} shows the error curves w/ and w/o weight prediction.
As shown in the figure, the RMSE of the predicted weights is apparently lower than the RMSE of the stale ones, indicating that the smoothed gradient based weight prediction is accurate and can help to mitigate the staleness issue.
As the version difference increases, the RMSE of the stale weights also increases, while the RMSE of the predicted weights is always much lower than the RMSE of the stale ones.

\section{Experiments}
\label{sec:exp}

\subsection{Evaluation Methodology} \label{subsec:exp_setup}

We use Falconwitch PS1816~\cite{h3}, a single-node multi-GPU system designed by H3 Platform, to conduct our performance and robustness studies.
The multi-GPU platform is equipped with four PCIe 3.0 x16 connected NVIDIA Tesla P40 GPUs, and it supports simultaneous execution of multiple peer-to-peer (P2P) transfers as long as the source and destination devices of these transfer requests are different. 
The CPU on the platform is Intel Xeon E5-2650 with 160GB DDR4-2400 off-chip main memory.

To evaluate the performance of the proposed method, six representative deep learning models are chosen as the benchmark.
These models cover different types of neural networks, including convolutional neural network (CNN), fully-connected network (FCN), and recurrent neural network (RNN).
For all these models, the mini-batch size is set to 128 and the momentum factor $\gamma$ is set to 0.9.
Since we focus on comparing the performance of different parallelization approaches rather than pursuing optimal accuracy, we adopt a fixed learning rate instead of variable rates to maintain the stability of the experiments. 

\paragraph{CNN Models}     
We use three state-of-the-art image classification CNNs, including \textbf{\emph{VGG16}}~\cite{vgg}, \textbf{\emph{ResNet-152}}~\cite{residualnet} and \textbf{\emph{Inception v4}}~\cite{inceptionv4}, to evaluate the impact of different parallelization methods on CNN models.
To train these models, we use CIFAR-10, which contains 50,000 training images and 10,000 testing images, as the dataset.
These three CNN models cover a wide range of layer shapes and sizes, and capture the evolution of CNNs in recent years.
We train these three models, \emph{VGG16}, \emph{ResNet-152}, and \emph{Inception v4}, using Momentum SGD with learning rates 2e-3, 2e-2, and 1e-3, respectively. Unlike the other two models, the first stage in \emph{VGG16} takes much longer execution time than other stages in the pipeline and thus becomes the bottleneck.  As a limitation of pipelining, this layer cannot be further divided for better load balance. Therefore, we combine pipeline with data parallelism and adopt data parallelism in the first stage to mitigate the imbalance between pipeline stages.

\paragraph{FCN Models}
For FCN models, we use \textbf{\emph{SNN}}~\cite{snn} (trained by CIFAR-10) and \textbf{\emph{Transformer}} (trained by IMDb Movie Review Sentiment Dataset~\cite{imdb}) to study the impact of different parallelization approaches.
\emph{SNN} is stacked by 32 fully-connected layers with 2048 hidden units, while \emph{Transformer} has 6 blocks in both encoder and decoder with 8 heads and 512 hidden units in fully-connected layers.
IMDb Movie Review Sentiment Dataset consists of 25,000 training movie reviews and 25,000 testing reviews, and each of the review is tagged with a sentiment, either positive or negative.
\emph{Transformer} is trained to determine the sentiment given input sentences.
Most of the parameters remain the same as the original model, except that the input sentences are truncated to 20 words.
These two models, \emph{SNN} and \emph{Transformer}, are trained using Momentum SGD with learning rates 1e-3 and 5e-5 respectively.

\paragraph{RNN Models}
We study the impact of different parallelization methods on \textbf{\emph{Residual LSTM}} \cite{residuallstm} using IMDb Dataset.
The \emph{Residual LSTM} comprises 8 layers of long short-term memory (LSTM) with 512 embedding units, 512 output units, and 1024 memory units.
During the training process, the learning rate is set to 5e-3.

We use TensorFlow~\cite{tensorflow} framework with a deep learning library, cuDNN~\cite{cudnn}, to implement these targeted neural network models.
TensorFlow provides high-level APIs for building data parallelism training process, and it also provides flexibility to implement model parallelism paradigms, including PipeDream and our proposed method.
We evaluate the following training implementations: 

\begin{itemize}
\item \textbf{Single GPU}: Training on single GPU. No parallelization method is applied.
\item \textbf{Data P.}: Data parallelism approach.
\item \textbf{Vanilla Model P.}: Model parallelism with pipeline.
\item \textbf{PipeDream}: Model parallelism with pipeline, and the staleness issue is mitigated by maintaining the consistency between the forward and backward tasks of the same GPU (Weight Stashing).
\item \textbf{SpecTrain}: Pipelined model parallelism with our proposed weight prediction mechanism (SpecTrain) to solve the staleness issue. We partition the model based on the same approach used by PipeDream.
\end{itemize}

\subsection{Throughput} \label{subsec:throughput}

\begin{figure*}[h]
    \includegraphics[width=.85\textwidth]{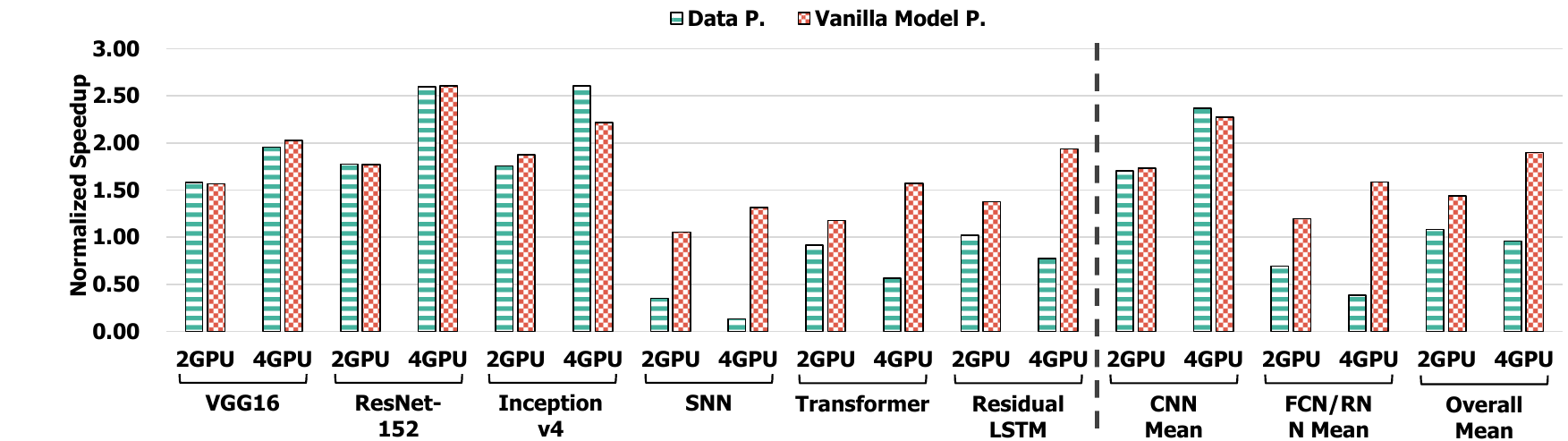}
    \caption{Throughput of the targeted neural network models with various parallelization approaches at 2-GPU and 4-GPU systems normalized to the throughput of Single GPU.}
    \label{fig:speedup}
\end{figure*}

Figure~\ref{fig:speedup} compares the throughput (measured as the number of training samples per second) for data and pipelined model parallelism. The throughput measurements are conducted over the interval between the 50th and 250th training iterations to get stable results. Reported throughput numbers are normalized to Single GPU.
Since the staleness mitigation method in PipeDream and SpecTrain have very little impact on performance, we only show the result of Vanilla Model P.

For CNN models, including \emph{VGG16}, \emph{ResNet-152}, and \emph{Inception v4}, the throughput of both parallel schemes increases as the number of GPUs increases. 
The throughput of pipelining is in par with data parallelism, except for \emph{Inception v4}, which contains less weights and thus imposes less communication overheads on data parallelism.

For FCN/RNN models, including \emph{SNN}, \emph{Transformer}, and \emph{Residual LSTM}, the scalability trend is different.
The 4-GPU results of Data P. show significant throughput degradation compared to the 2-GPU results for all these three models.
On average, at the 4-GPU system, Data P. can only provide 38.5\% throughput improvement compared to Single GPU.
The reason is that FCN and RNN models typically contain a rich amount of weights, and, for each mini-batch, these weights should be synchronized among all GPUs when Data P. is applied.
Detailed analysis about the synchronization overhead incurred by Data P. will be discussed in Section~\ref{subsec:perf_breakdown}.
Different from Data P., the model parallelism scheme gets significant throughput improvement on the 4-GPU system, as they do not suffer from the huge synchronization overhead.
On average, when running FCN/RNN models on the 4-GPU system, model parallelism provides 3.10x throughput improvement over Data P..

In summary, model parallelism provides a pipeline design without incurring throughput-harmful synchronization overheads.
On average, at the 4-GPU system, model parallelism provides 98.5\% throughput improvement (up to 8.91x) over Data P.

\subsection{Performance Breakdown} \label{subsec:perf_breakdown}

\begin{figure*}[h]
    \centering
    \includegraphics[width=.9\textwidth]{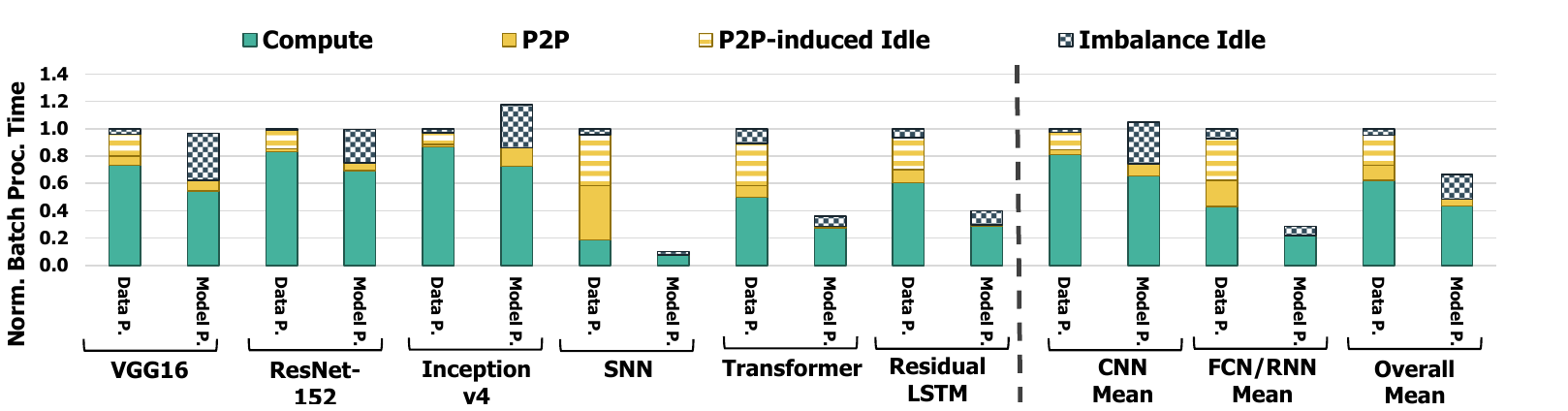}
    \caption{Performance breakdown of Data P. and Vanilla Model P. normalized to Data P..}
    \label{fig:breakdown}
\end{figure*}

\begin{figure*}[h]
    \centering
    \subfigure[Train Loss] {
        \centering
        \includegraphics[width=0.3\textwidth]{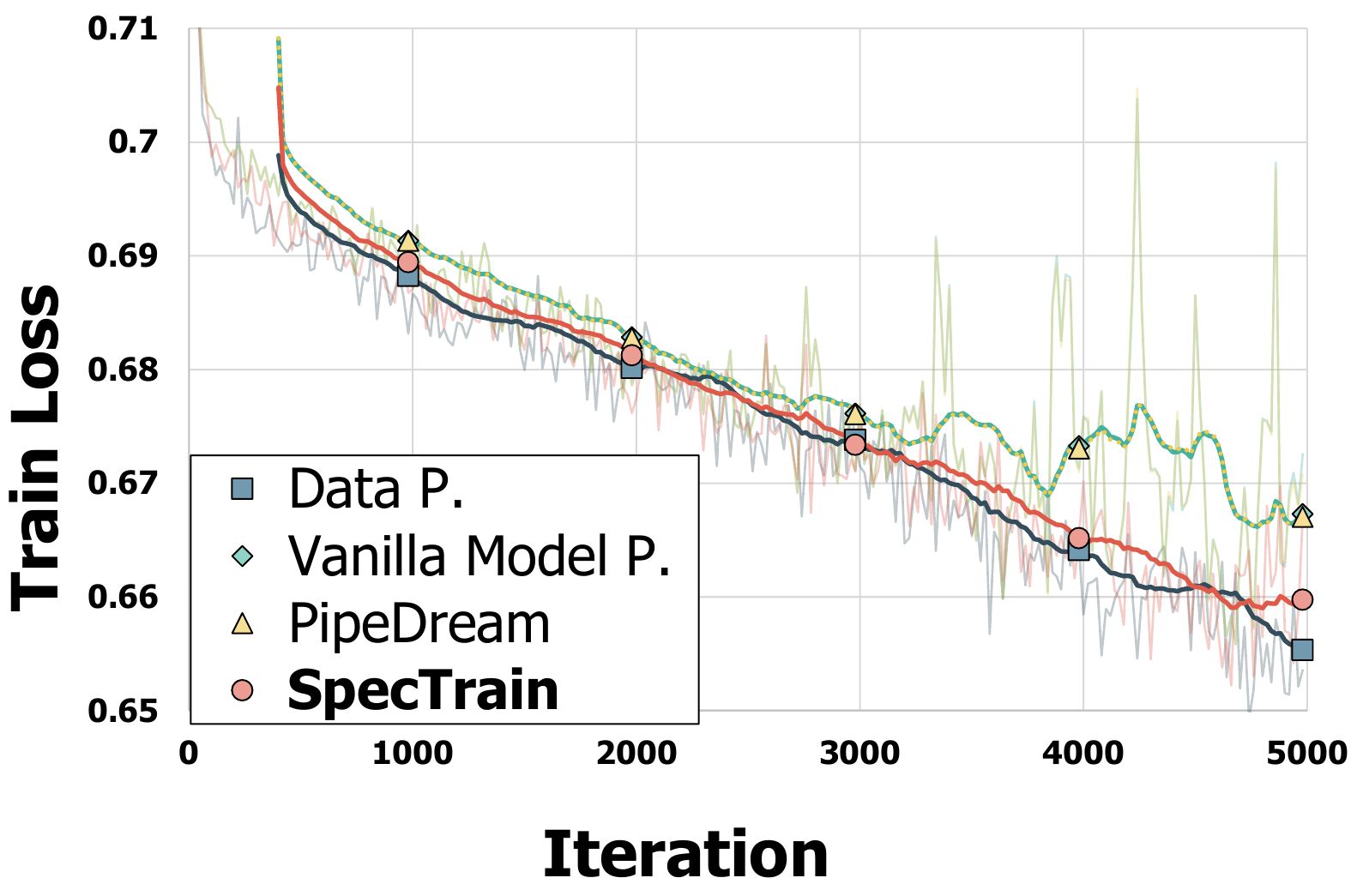}
    }
    \subfigure[Validation Loss] {
        \centering
        \includegraphics[width=0.3\textwidth]{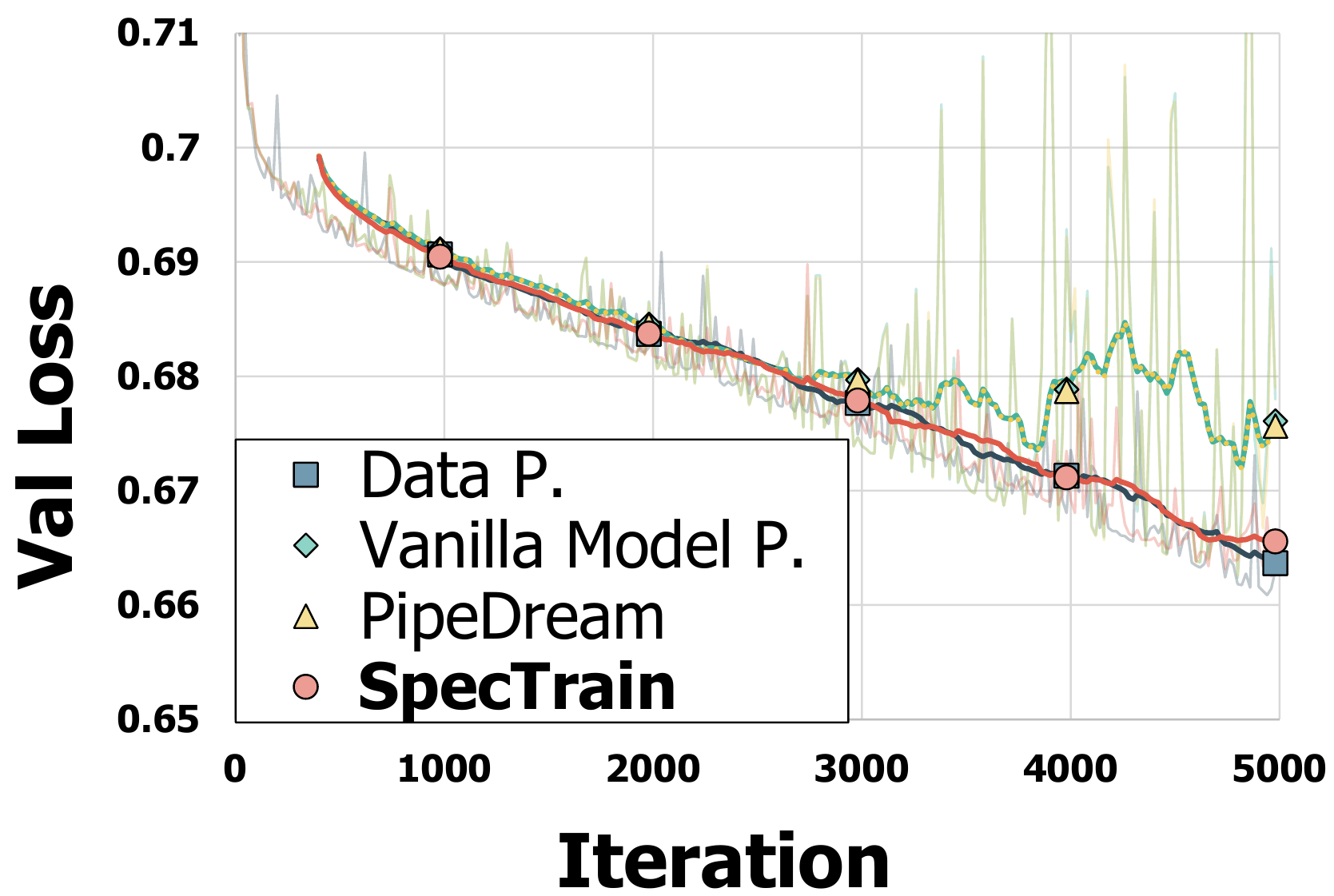}
    }
    \subfigure[Validation Accuracy] {
        \centering
        \includegraphics[width=0.31\textwidth]{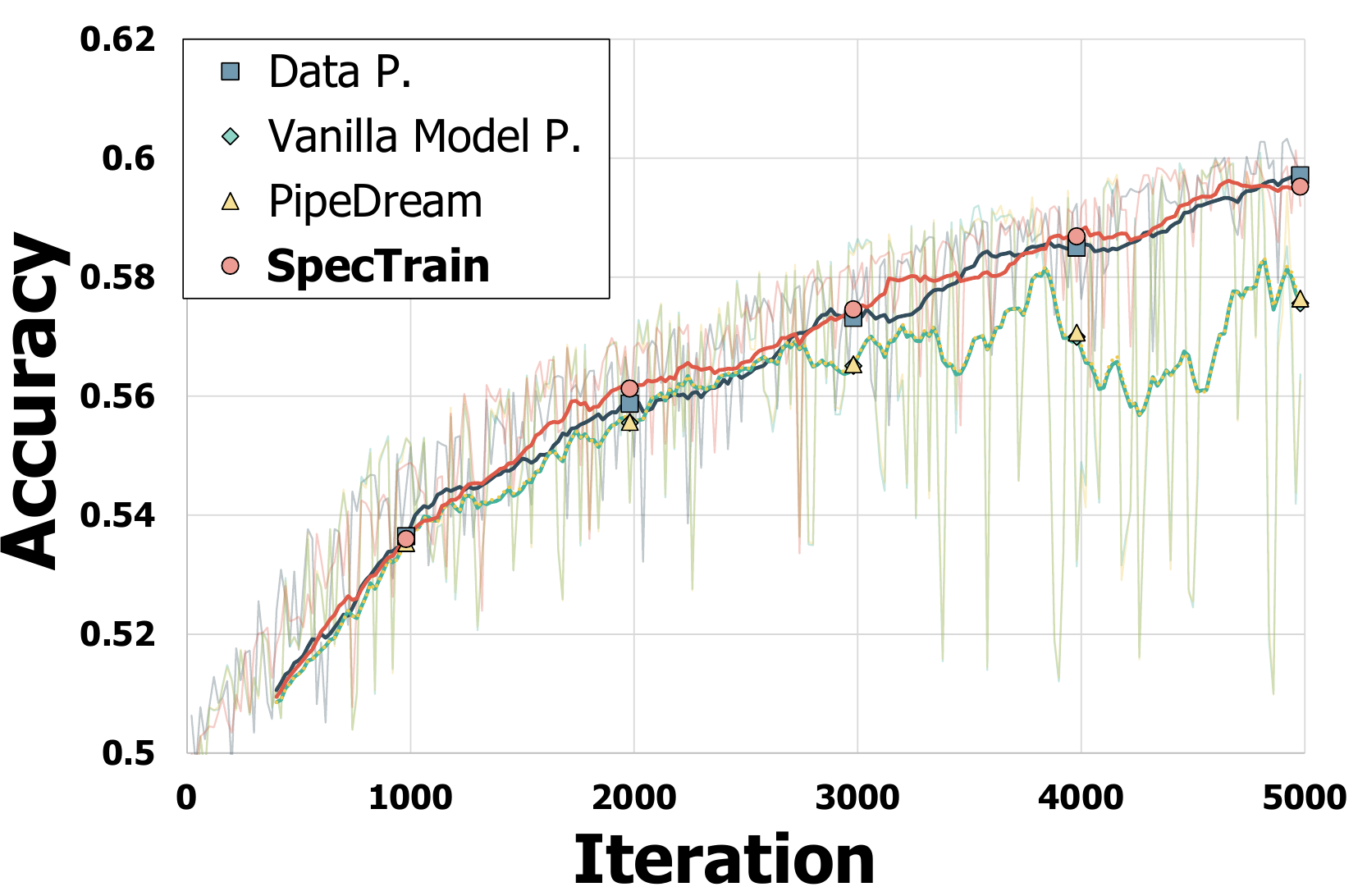}
    }
    \caption{Learning curves when using different parallelization schemes to train \emph{Transformer}. The light-color lines denote the measured values. To make the trends clear, we add a thick line to show the moving average over 20 latest values.}
    \label{fig:transformer_curve}
\end{figure*}

To analyze the performance difference between data parallelism and model parallelism in detail, we breakdown the overall execution time into \emph{computing time}, \emph{P2P transfer time}, \emph{P2P-induced idle time}, and \emph{imbalance-induced idle time}. The computing time includes the time spent on computing. P2P transfer time represents the data transfer time that cannot be overlapped with computing, and P2P-induced idle time is the time that data transfer is stalled due to busy PCI-E interconnection. Imbalance-induced idle is the GPU idle time waiting for other GPUs to come to the synchronization point, i.e., imbalanced workloads among GPUs. 

We use NVIDIA CUDA 8.0 profiler (nvprof) to get the runtime data related to performance breakdown during the training process.
First, we dump the GPU trace by executing the training program with nvprof on the 4-GPU system.
The trace contains the starting timestamps, ending timestamps, and transfer size, of every computing kernel and data transfer transaction.
We then write a script to analyze the GPU trace and get the performance breakdown.

Figure~\ref{fig:breakdown} illustrates the performance breakdown of different parallelization methods on the 4-GPU system when processing a mini-batch, normalized to the overall execution time of a mini-batch when Data P. is applied.
Data P. spends more time on P2P transfer and P2P-induced idle than the model parallelism scheme, as Data P. transfers huge amount of data during weight synchronization, especially when training FCN/RNN models.
On average, the P2P-related time (i.e., P2P transfer and P2P-induced idle) accounts for 26.7\% overall execution time when Data P. is applied, while the P2P-related time in pipelined model parallelism is only 4.52\%.
For FCN/RNN models, the P2P-related time in Data P. contributes to 49.8\% of total execution time on average, indicating that the communication among GPUs would greatly impact performance. In the aspect of workload balance among GPUs, model parallelism shows higher imbalance-induced idle time than data parallelism.  


The computing time of Data P. is 63.1\% longer than model parallelism on average.
One reason is that the mini-batch partition leads to GPU under-utilization.
Another reason is due to kernel preprocessing recomputation. 
The state-of-the-art algorithms for computation in neural networks, such as FFT \cite{fft} and Winograd \cite{winograd}, require kernels to preprocess weights before computing the input data. 
This means that, for data parallelism, every GPU needs to reperform the preprocessing on the replicated weights, thus increasing the computational overheads.

\subsection{Staleness and Convergence} \label{subsec:stale_conv}

\begin{table}[t]
\small
\centering
\caption{Model accuracy of different neural network models when various parallelization schemes are applied. The bold values indicates the best one among the three pipelining implementations.}
\begin{tabular}{rccc}
    \toprule
    \makecell{Parallelization \\ Scheme} & \makecell{Min. \\ Train Loss} & \makecell{Min. \\ Val. Loss} & \makecell {Max. Val. \\ Accuracy} \\
    \midrule
    \makecell[l]{\textbf{VGG16}} \\
    \cmidrule(lr){1-1}
    \textit{Data P.} & \textit{0.213271} & \textit{0.794613} & \textit{73.4776\%} \\
    Vanilla Model P. & 0.204126 & 0.811148 & 73.0569\% \\
    PipeDream & 0.200585 & 0.811144 & 72.8365\% \\
    SpecTrain & \textbf{0.185566} & \textbf{0.796017} & \textbf{73.8081\%} \\
    \midrule
    \makecell[l]{\textbf{ResNet-152}} \\
    \cmidrule(lr){1-1}
    \textit{Data P.} & \textit{0.338845} & \textit{0.892366} & \textit{71.2139\%} \\
    Vanilla Model P. & 0.254327 & 0.945588 & 70.3225\% \\
    PipeDream & 0.467527 & 0.979401 & 67.9287\% \\
    SpecTrain & \textbf{0.231724} & \textbf{0.924241} & \textbf{70.9936\%} \\
    \midrule
    \makecell[l]{\textbf{Inception v4}} \\
    \cmidrule(lr){1-1}
    \textit{Data P.} & \textit{0.804475} & \textit{0.913155} & \textit{69.1607\%} \\
    Vanilla Model P. & 0.858834 & 0.919470 & 68.6599\% \\
    PipeDream & 0.864199 & 0.930320 & 68.5297\% \\
    SpecTrain & \textbf{0.756939} & \textbf{0.874898} & \textbf{70.7732\%} \\
    \midrule
    \makecell[l]{\textbf{SNN}} \\
    \cmidrule(lr){1-1}
    \textit{Data P.} & \textit{0.431832} & \textit{1.45124} & \textit{50.9115\%} \\
    Vanilla Model P. & 0.766552 & 1.440200 & 50.6911\% \\
    PipeDream & 0.810452 & 1.450239 & 50.4107\% \\
    SpecTrain & \textbf{0.724402} & \textbf{1.406447} & \textbf{52.0733\%} \\
    \midrule
    \makecell[l]{\textbf{Transformer}} \\
    \cmidrule(lr){1-1}
    \textit{Data P.} & \textit{0.649287} & \textit{0.660871} & \textit{60.3265\%} \\
    Vanilla Model P. & 0.655801 & \textbf{0.662379} & 60.0963\% \\
    PipeDream & 0.655877 & 0.662544 & 59.9760\% \\
    SpecTrain & \textbf{0.652193} & 0.662502 & \textbf{60.1362\%} \\
    \midrule
    \makecell[l]{\textbf{Residual LSTM}} \\
    \cmidrule(lr){1-1}
    \textit{Data P.} & \textit{0.347742} & \textit{0.658583} & \textit{66.0557\%} \\
    Vanilla Model P. & 0.459975 & 0.652651 & \textbf{65.0240\%} \\
    PipeDream & 0.467595 & 0.652948 & 64.8137\% \\
    SpecTrain & \textbf{0.454813} & \textbf{0.652251} & 64.8137\% \\
    \bottomrule
\end{tabular}
\label{tab:convergence}
\end{table}

Since pipelined model parallelism induces the staleness problem, we analyze the impact of staleness on convergence for different model parallelism based parallelization approaches, including Vanilla Model P., PipeDream, and the SpecTrain we proposed.
These model parallelism based parallelization methods are also compared with Data P., which has no staleness issues.
Vanilla Model P. represents the worst case, as no staleness mitigation technique is applied.
PipeDream mitigates the inconsistency between the forward and backward pass by sticking to stale weights while SpecTrain utilizes weight prediction to alleviate the staleness issue.
The experiments are done by training each model for 5000 iterations.
The learning rates of these models are tuned, as described in Section~\ref{subsec:exp_setup}, to ensure the training converges eventually.
Training loss, validation loss, and validation accuracy are recorded every 20 steps (iterations) to show the learning curve. 
In the following paragraphs, we analyze the impact of staleness problem on learning curve and model accuracy.

Figure~\ref{fig:transformer_curve} shows the learning curve of \emph{Transformer} when different staleness mitigation techniques are applied.
Due to the space limitation, only the learning curve of \emph{Transformer} is illustrated, but the trend is similar for all neural network models.
As shown in Figure~\ref{fig:transformer_curve}, Vanilla Model P. is the least stable parallelization scheme, especially near the convergence point, as no staleness mitigation technique is applied.
PipeDream mitigates the weight inconsistency issue between the forward and backward pass but still uses stale weights, so the improvement is limited and the learning curve is very close to Vanilla Model P.
Our SpecTrain technique effectively alleviates the instability problem by weight prediction and the learning curve of SpecTrain is similar to Data P., indicating that using SpecTrain can achieve near robust training process. 

Table~\ref{tab:convergence} shows the results of model accuracy.
Vanilla Model P., the Model P. without staleness mitigation technique, is seriously affected by the staleness problem, resulting in 0.55\% accuracy drop on average, compared to Data P..
The staleness mitigation method adopted by PipeDream cannot alleviate the staleness issue in our experiments, losing 1.1\% validation accuracy over Data P. on average, which is worse than Vanilla Model P.
It is likely because the adoption of the stale weights in forward pass aggravates staleness, although it resolves the weight inconsistency issue.
By using weight prediction to resolve the staleness issue, our SpecTrain shows no accuracy drop in almost all of the workloads and achieves the same validation accuracy as the staleness-free Data P.
The only exception is the RNN model, \emph{Residual LSTM}, most likely due to the gradient exploding issue~\cite{DBLP:conf/icml/PascanuMB13}.
The gradients in RNNs sometimes dramatically increase, and a sudden exploding gradient lowers the prediction accuracy of SpecTrain. 

\section{Related Work}
\label{sec:relatedwork}

Most of related studies regarding DNN training parallelization target on distributed systems.
Data parallalism is the most commonly used parallelization scheme in distributed systems.
Many deep learning frameworks such as Tensorflow \cite{tensorflow}, PyTorch \cite{pytorch} and Caffe \cite{caffe} provide high-level APIs for applying data parallel training.
It is well known that the huge communication overhead of data parallelism limits its scalability.
Some of prior studies focus on reducing the communication overheads for data parallelism.
Seide et al. \cite{DBLP:conf/interspeech/SeideFDLY14} propose 1-bit SGD, which quantizes the gradients to one bit per value to lower the communication cost.
Strom \cite{DBLP:conf/interspeech/Strom15} proposes to send gradients only when it is larger than a threshold to reduce communications.
Alistarh et al. \cite{DBLP:journals/corr/Alistarh0TV16} design another gradient quantization approach to do compression, and their proposed scheme allows users to trade off accuracy and performance.
Zhou et al. \cite{DBLP:journals/corr/ZhouNZWWZ16} speed up data parallelism on convolutional neural networks by using low-bit-width weights, activations, and gradients.
Aji et al. \cite{DBLP:conf/emnlp/AjiH17} propose Gradient Dropping to prevent frequent gradient transfers.
Lin et al. \cite{DBLP:journals/corr/abs-1712-01887} implement Gradient Dropping \cite{DBLP:conf/emnlp/AjiH17}, and develop many techniques to increase the robustness.
Sun et al. \cite{DBLP:conf/ispan/SunZYZBL17} improve the performance of parameter server by leveraging distributed shared memory to reduce the networking time.
Mamidala et al. \cite{DBLP:journals/corr/abs-1801-03855} apply MPI parallelism to replace the gradient aggregation process at the parameter server, in order to provide better scalability.
Project Adam \cite{DBLP:conf/osdi/ChilimbiSAK14} and FireCaffe \cite{DBLP:conf/cvpr/IandolaMAK16} design deep learning training systems that optimize and balance workload computation and communication to scale DNN. 


In distributed systems, another issue, straggler problem, further downgrades the performance of data parallelism.
Due to unstable networking, some workers may encounter heavy traffic and consume much time on weight synchronization, making other workers stall.
This straggler problem reduces GPU utilization.
Chen et al. \cite{DBLP:journals/corr/ChenMBJ16} propose to give up slow workers at run-time, solving the straggler problem without affecting algorithmic correctness. 
Dean et al. \cite{DBLP:conf/nips/DeanCMCDLMRSTYN12} propose a novel variant of SGD called asynchronous data parallelism, which skips the synchronization between workers.
Some deep learning frameworks like Tensorflow \cite{tensorflow} and Theano-MPI \cite{Theano-mpi} have supported this parallelization approach.
However, without synchronization, workers may use stale weights to calculate.
As such, asynchronous data parallelism also faces the staleness issue.
Cipar et al. \cite{DBLP:conf/hotos/CiparHKLGGKX13} present bounded staleness to prevent calculating gradients based on excessively stale weights.
Zhang et al. \cite{DBLP:conf/ijcai/ZhangGL016} propose staleness-dependant learning rate to apply stale gradients on weight updates with a decay proportional to the inverse of the weight version difference.



Model parallelism is another appealing approach to parallelize DNN training.
Krizhevsky \cite{DBLP:journals/corr/Krizhevsky14} proposes to apply model parallelism on fully-connected layers of AlexNet \cite{alexnet}.
Li et al.\cite{DBLP:conf/ijcnn/LiZHD0XZY14} design a two-stage pipeline for training recurrent neural network.
Mirhoseini et al. \cite{DBLP:conf/icml/MirhoseiniPLSLZ17} use reinforcement learning to come up with a device placement policy for model parallelism.
The above methods cannot parallelize some portions of a model if dependencies exist, i.e. inter-layer parallelization.
Harlap et al. propose PipeDream \cite{pipedream} as the first work using pipeline in model parallelism to train DNN models, enabling inter-layer parallelization. 
Huo et al. \cite{DBLP:conf/icml/HuoGYH18,DBLP:journals/corr/abs-1807-04511} also propose algorithms to break backward locking for model parallelism so that the backward pass can be performed in parallel. GPipe \cite{DBLP:journals/corr/abs-1811-06965} updates the weights synchronously and reduce the GPU idle time by dividing the mini-batch into multiple micro-batches and piplining the micro-batches divided from the same mini-batch.

Some recent works consider not only data parallelism and model parallelism when finding parallelization strategy. Jia et al. \cite{DBLP:journals/corr/abs-1807-05358, DBLP:conf/icml/JiaLQA18} partition the tensor along multiple dimensions and search for best parallelization strategy.

\section{Conclusion}
\label{sec:conclusion}

As training DNN models is time-consuming, multi-GPU acceleration is a widely adopted way to speed up.
Data parallelism is the most common parallelization scheme but it suffers from huge inter-GPU communication overheads.
Model parallelism provides another parallelization method with much less communications, however, posing a challenge: staleness issue.
We propose a weight prediction method, SpecTrain, that leverages smoothed gradients for weight prediction to mitigate the staleness problem.
Experimental results show that the throughput of pipelined model parallelism is 98.5\% higher than data parallelism on average (up to 8.91x).
In addition, SpecTrain shows robustness on training with no accuracy drop in most of the workloads compared to staleness-free training.



\printAffiliationsAndNotice{}

\nocite{langley00}

\bibliography{references}

\begin{thebibliography}{50}
\providecommand{\natexlab}[1]{#1}
\providecommand{\url}[1]{\texttt{#1}}
\expandafter\ifx\csname urlstyle\endcsname\relax
  \providecommand{\doi}[1]{doi: #1}\else
  \providecommand{\doi}{doi: \begingroup \urlstyle{rm}\Url}\fi

\bibitem[Abadi et~al.(2015)Abadi, Agarwal, Barham, Brevdo, Chen, Citro,
  Corrado, Davis, Dean, Devin, Ghemawat, Goodfellow, Harp, Irving, Isard, Jia,
  Jozefowicz, Kaiser, Kudlur, Levenberg, Man\'{e}, Monga, Moore, Murray, Olah,
  Schuster, Shlens, Steiner, Sutskever, Talwar, Tucker, Vanhoucke, Vasudevan,
  Vi\'{e}gas, Vinyals, Warden, Wattenberg, Wicke, Yu, and Zheng]{tensorflow}
Abadi, M., Agarwal, A., Barham, P., Brevdo, E., Chen, Z., Citro, C., Corrado,
  G.~S., Davis, A., Dean, J., Devin, M., Ghemawat, S., Goodfellow, I., Harp,
  A., Irving, G., Isard, M., Jia, Y., Jozefowicz, R., Kaiser, L., Kudlur, M.,
  Levenberg, J., Man\'{e}, D., Monga, R., Moore, S., Murray, D., Olah, C.,
  Schuster, M., Shlens, J., Steiner, B., Sutskever, I., Talwar, K., Tucker, P.,
  Vanhoucke, V., Vasudevan, V., Vi\'{e}gas, F., Vinyals, O., Warden, P.,
  Wattenberg, M., Wicke, M., Yu, Y., and Zheng, X.
\newblock {TensorFlow}: Large-scale machine learning on heterogeneous systems,
  2015.
\newblock URL \url{https://www.tensorflow.org/}.

\bibitem[Aji \& Heafield(2017)Aji and Heafield]{DBLP:conf/emnlp/AjiH17}
Aji, A.~F. and Heafield, K.
\newblock Sparse communication for distributed gradient descent.
\newblock In \emph{Proceedings of the Conference on Empirical Methods in
  Natural Language Processing}, pp.\  440--445, 2017.

\bibitem[Alistarh et~al.(2016)Alistarh, Li, Tomioka, and
  Vojnovic]{DBLP:journals/corr/Alistarh0TV16}
Alistarh, D., Li, J., Tomioka, R., and Vojnovic, M.
\newblock {QSGD:} randomized quantization for communication-optimal stochastic
  gradient descent.
\newblock \emph{CoRR}, abs/1610.02132, 2016.
\newblock URL \url{http://arxiv.org/abs/1610.02132}.

\bibitem[Arivazhagan et~al.(2019)Arivazhagan, Bapna, Firat, Lepikhin, Johnson,
  Krikun, Chen, Cao, Foster, Cherry, Macherey, Chen, and
  Wu]{DBLP:journals/corr/abs-1907-05019}
Arivazhagan, N., Bapna, A., Firat, O., Lepikhin, D., Johnson, M., Krikun, M.,
  Chen, M.~X., Cao, Y., Foster, G., Cherry, C., Macherey, W., Chen, Z., and Wu,
  Y.
\newblock Massively multilingual neural machine translation in the wild:
  Findings and challenges.
\newblock \emph{CoRR}, abs/1907.05019, 2019.

\bibitem[Bahdanau et~al.(2014)Bahdanau, Cho, and
  Bengio]{DBLP:journals/corr/BahdanauCB14}
Bahdanau, D., Cho, K., and Bengio, Y.
\newblock Neural machine translation by jointly learning to align and
  translate.
\newblock \emph{CoRR}, abs/1409.0473, 2014.
\newblock URL \url{http://arxiv.org/abs/1409.0473}.

\bibitem[Chen et~al.(2016)Chen, Monga, Bengio, and
  J{\'{o}}zefowicz]{DBLP:journals/corr/ChenMBJ16}
Chen, J., Monga, R., Bengio, S., and J{\'{o}}zefowicz, R.
\newblock Revisiting distributed synchronous {SGD}.
\newblock \emph{CoRR}, abs/1604.00981, 2016.
\newblock URL \url{http://arxiv.org/abs/1604.00981}.

\bibitem[Chetlur et~al.(2014)Chetlur, Woolley, Vandermersch, Cohen, Tran,
  Catanzaro, and Shelhamer]{cudnn}
Chetlur, S., Woolley, C., Vandermersch, P., Cohen, J., Tran, J., Catanzaro, B.,
  and Shelhamer, E.
\newblock cudnn: Efficient primitives for deep learning.
\newblock \emph{CoRR}, abs/1410.0759, 2014.
\newblock URL \url{http://arxiv.org/abs/1410.0759}.

\bibitem[Chilimbi et~al.(2014)Chilimbi, Suzue, Apacible, and
  Kalyanaraman]{DBLP:conf/osdi/ChilimbiSAK14}
Chilimbi, T.~M., Suzue, Y., Apacible, J., and Kalyanaraman, K.
\newblock Project adam: Building an efficient and scalable deep learning
  training system.
\newblock In \emph{Proceedings of the 11th {USENIX} Symposium on Operating
  Systems Design and Implementation}, pp.\  571--582, 2014.

\bibitem[Cipar et~al.(2013)Cipar, Ho, Kim, Lee, Ganger, Gibson, Keeton, and
  Xing]{DBLP:conf/hotos/CiparHKLGGKX13}
Cipar, J., Ho, Q., Kim, J.~K., Lee, S., Ganger, G.~R., Gibson, G., Keeton, K.,
  and Xing, E.~P.
\newblock Solving the straggler problem with bounded staleness.
\newblock In \emph{Proceedings of the 14th Workshop on Hot Topics in Operating
  Systems}, 2013.

\bibitem[Dean et~al.(2012)Dean, Corrado, Monga, Chen, Devin, Le, Mao, Ranzato,
  Senior, Tucker, Yang, and Ng]{DBLP:conf/nips/DeanCMCDLMRSTYN12}
Dean, J., Corrado, G., Monga, R., Chen, K., Devin, M., Le, Q.~V., Mao, M.~Z.,
  Ranzato, M., Senior, A.~W., Tucker, P.~A., Yang, K., and Ng, A.~Y.
\newblock Large scale distributed deep networks.
\newblock In \emph{Proceedings of the 26th Conference on Neural Information
  Processing Systems}, pp.\  1232--1240, 2012.

\bibitem[{H3 Platform Inc.}()]{h3}
{H3 Platform Inc.}
\newblock Falconwitch ps1816.
\newblock \url{http://www.h3platform.com/product/}.

\bibitem[Harlap(2019)]{pipedream_thesis}
Harlap, A.
\newblock Improving ml applications in shared computing environments.
\newblock \emph{Doctoral dissertation}, 2019.
\newblock URL
  \url{https://aaronharlap.github.io//papers/aharlap_dissertation.pdf}.

\bibitem[Harlap et~al.(2018)Harlap, Narayanan, Phanishayee, Seshadri, Devanur,
  Ganger, and Gibbons]{pipedream}
Harlap, A., Narayanan, D., Phanishayee, A., Seshadri, V., Devanur, N.~R.,
  Ganger, G.~R., and Gibbons, P.~B.
\newblock Pipedream: Fast and efficient pipeline parallel {DNN} training.
\newblock \emph{CoRR}, abs/1806.03377, 2018.
\newblock URL \url{http://arxiv.org/abs/1806.03377}.

\bibitem[He et~al.(2015)He, Zhang, Ren, and Sun]{residualnet}
He, K., Zhang, X., Ren, S., and Sun, J.
\newblock Deep residual learning for image recognition.
\newblock \emph{CoRR}, abs/1512.03385, 2015.
\newblock URL \url{http://arxiv.org/abs/1512.03385}.

\bibitem[Hinton et~al.(2012)Hinton, Deng, Yu, Dahl, Mohamed, Jaitly, Senior,
  Vanhoucke, Nguyen, Sainath, et~al.]{hinton2012deep}
Hinton, G., Deng, L., Yu, D., Dahl, G.~E., Mohamed, A.-r., Jaitly, N., Senior,
  A., Vanhoucke, V., Nguyen, P., Sainath, T.~N., et~al.
\newblock Deep neural networks for acoustic modeling in speech recognition: The
  shared views of four research groups.
\newblock \emph{IEEE Signal processing magazine}, 29\penalty0 (6):\penalty0
  82--97, 2012.

\bibitem[Huang et~al.(2018)Huang, Cheng, Chen, Lee, Ngiam, Le, and
  Chen]{DBLP:journals/corr/abs-1811-06965}
Huang, Y., Cheng, Y., Chen, D., Lee, H., Ngiam, J., Le, Q.~V., and Chen, Z.
\newblock Gpipe: Efficient training of giant neural networks using pipeline
  parallelism.
\newblock \emph{CoRR}, abs/1811.06965, 2018.
\newblock URL \url{http://arxiv.org/abs/1811.06965}.

\bibitem[Huo et~al.(2018{\natexlab{a}})Huo, Gu, and
  Huang]{DBLP:journals/corr/abs-1807-04511}
Huo, Z., Gu, B., and Huang, H.
\newblock Training neural networks using features replay.
\newblock \emph{CoRR}, abs/1807.04511, 2018{\natexlab{a}}.
\newblock URL \url{http://arxiv.org/abs/1807.04511}.

\bibitem[Huo et~al.(2018{\natexlab{b}})Huo, Gu, Yang, and
  Huang]{DBLP:conf/icml/HuoGYH18}
Huo, Z., Gu, B., Yang, Q., and Huang, H.
\newblock Decoupled parallel backpropagation with convergence guarantee.
\newblock In \emph{Proceedings of the 35th International Conference on Machine
  Learning}, pp.\  2103--2111, 2018{\natexlab{b}}.

\bibitem[Iandola et~al.(2016)Iandola, Moskewicz, Ashraf, and
  Keutzer]{DBLP:conf/cvpr/IandolaMAK16}
Iandola, F.~N., Moskewicz, M.~W., Ashraf, K., and Keutzer, K.
\newblock Firecaffe: Near-linear acceleration of deep neural network training
  on compute clusters.
\newblock In \emph{Proceedings of the Conference on Computer Vision and Pattern
  Recognition}, pp.\  2592--2600, 2016.

\bibitem[Jia et~al.(2014)Jia, Shelhamer, Donahue, Karayev, Long, Girshick,
  Guadarrama, and Darrell]{caffe}
Jia, Y., Shelhamer, E., Donahue, J., Karayev, S., Long, J., Girshick, R.,
  Guadarrama, S., and Darrell, T.
\newblock Caffe: Convolutional architecture for fast feature embedding.
\newblock \emph{CoRR}, 2014.
\newblock URL \url{https://arxiv.org/abs/1408.5093}.

\bibitem[Jia et~al.(2018)Jia, Lin, Qi, and Aiken]{DBLP:conf/icml/JiaLQA18}
Jia, Z., Lin, S., Qi, C.~R., and Aiken, A.
\newblock Exploring hidden dimensions in parallelizing convolutional neural
  networks.
\newblock In \emph{Proceedings of the 35th International Conference on Machine
  Learning}, pp.\  2279--2288, 2018.

\bibitem[Jia et~al.(2019)Jia, Zaharia, and
  Aiken]{DBLP:journals/corr/abs-1807-05358}
Jia, Z., Zaharia, M., and Aiken, A.
\newblock Beyond data and model parallelism for deep neural networks.
\newblock In \emph{Proceedings of the 2nd Conference on Systems and Machine
  Learning}, 2019.

\bibitem[Keskar et~al.(2016)Keskar, Mudigere, Nocedal, Smelyanskiy, and
  Tang]{DBLP:journals/corr/KeskarMNST16}
Keskar, N.~S., Mudigere, D., Nocedal, J., Smelyanskiy, M., and Tang, P. T.~P.
\newblock On large-batch training for deep learning: Generalization gap and
  sharp minima.
\newblock \emph{CoRR}, abs/1609.04836, 2016.
\newblock URL \url{http://arxiv.org/abs/1609.04836}.

\bibitem[Kim et~al.(2017)Kim, El{-}Khamy, and Lee]{residuallstm}
Kim, J., El{-}Khamy, M., and Lee, J.
\newblock Residual {LSTM:} design of a deep recurrent architecture for distant
  speech recognition.
\newblock In \emph{Proceedings of the 18th Conference on International Speech
  Communication Association}, pp.\  1591--1595, 2017.

\bibitem[Kingma \& Ba(2014)Kingma and Ba]{adam}
Kingma, D.~P. and Ba, J.
\newblock Adam: {A} method for stochastic optimization.
\newblock \emph{CoRR}, abs/1412.6980, 2014.
\newblock URL \url{http://arxiv.org/abs/1412.6980}.

\bibitem[Klambauer et~al.(2017)Klambauer, Unterthiner, Mayr, and
  Hochreiter]{snn}
Klambauer, G., Unterthiner, T., Mayr, A., and Hochreiter, S.
\newblock Self-normalizing neural networks.
\newblock In \emph{Proceedings of the 31st Conference on Neural Information
  Processing Systems}, pp.\  972--981, 2017.

\bibitem[Krizhevsky(2014)]{DBLP:journals/corr/Krizhevsky14}
Krizhevsky, A.
\newblock One weird trick for parallelizing convolutional neural networks.
\newblock \emph{CoRR}, abs/1404.5997, 2014.
\newblock URL \url{http://arxiv.org/abs/1404.5997}.

\bibitem[Krizhevsky et~al.(2012)Krizhevsky, Sutskever, and Hinton]{alexnet}
Krizhevsky, A., Sutskever, I., and Hinton, G.~E.
\newblock Imagenet classification with deep convolutional neural networks.
\newblock In \emph{Proceedings of the 26th Conference on Neural Information
  Processing Systems}, pp.\  1106--1114, 2012.

\bibitem[Lavin \& Gray(2016)Lavin and Gray]{winograd}
Lavin, A. and Gray, S.
\newblock Fast algorithms for convolutional neural networks.
\newblock In \emph{Proceedings of the Conference on Computer Vision and Pattern
  Recognition}, pp.\  4013--4021, 2016.

\bibitem[Li et~al.(2014)Li, Zhou, Huang, Duan, Wang, Xu, Zhang, and
  Yang]{DBLP:conf/ijcnn/LiZHD0XZY14}
Li, B., Zhou, E., Huang, B., Duan, J., Wang, Y., Xu, N., Zhang, J., and Yang,
  H.
\newblock Large scale recurrent neural network on {GPU}.
\newblock In \emph{Proceedings of the International Joint Conference on Neural
  Networks}, pp.\  4062--4069, 2014.

\bibitem[Lian et~al.(2015)Lian, Huang, Li, and Liu]{DBLP:conf/nips/LianHLL15}
Lian, X., Huang, Y., Li, Y., and Liu, J.
\newblock Asynchronous parallel stochastic gradient for nonconvex optimization.
\newblock In \emph{Proceedings of the 29th Conference on Neural Information
  Processing Systems}, pp.\  2737--2745, 2015.

\bibitem[Lin et~al.(2017)Lin, Han, Mao, Wang, and
  Dally]{DBLP:journals/corr/abs-1712-01887}
Lin, Y., Han, S., Mao, H., Wang, Y., and Dally, W.~J.
\newblock Deep gradient compression: Reducing the communication bandwidth for
  distributed training.
\newblock \emph{CoRR}, abs/1712.01887, 2017.
\newblock URL \url{http://arxiv.org/abs/1712.01887}.

\bibitem[Ma et~al.(2016)Ma, Mao, and Taylor]{Theano-mpi}
Ma, H., Mao, F., and Taylor, G.~W.
\newblock Theano-mpi: {A} theano-based distributed training framework.
\newblock In \emph{Proceedings of the 22nd International European Conference on
  Parallel and Distributed Computing}, pp.\  800--813, 2016.

\bibitem[Maas et~al.(2011)Maas, Daly, Pham, Huang, Ng, and Potts]{imdb}
Maas, A.~L., Daly, R.~E., Pham, P.~T., Huang, D., Ng, A.~Y., and Potts, C.
\newblock Learning word vectors for sentiment analysis.
\newblock In \emph{Proceedings of the 49th Annual Meeting of the Association
  for Computational Linguistics}, pp.\  142--150, 2011.

\bibitem[Mamidala et~al.(2018)Mamidala, Kollias, Ward, and
  Artico]{DBLP:journals/corr/abs-1801-03855}
Mamidala, A.~R., Kollias, G., Ward, C., and Artico, F.
\newblock {MXNET-MPI:} embedding {MPI} parallelism in parameter server task
  model for scaling deep learning.
\newblock \emph{CoRR}, abs/1801.03855, 2018.
\newblock URL \url{http://arxiv.org/abs/1801.03855}.

\bibitem[Mathieu et~al.(2013)Mathieu, Henaff, and LeCun]{fft}
Mathieu, M., Henaff, M., and LeCun, Y.
\newblock Fast training of convolutional networks through ffts.
\newblock \emph{CoRR}, abs/1312.5851, 2013.
\newblock URL \url{http://arxiv.org/abs/1312.5851}.

\bibitem[Mirhoseini et~al.(2017)Mirhoseini, Pham, Le, Steiner, Larsen, Zhou,
  Kumar, Norouzi, Bengio, and Dean]{DBLP:conf/icml/MirhoseiniPLSLZ17}
Mirhoseini, A., Pham, H., Le, Q.~V., Steiner, B., Larsen, R., Zhou, Y., Kumar,
  N., Norouzi, M., Bengio, S., and Dean, J.
\newblock Device placement optimization with reinforcement learning.
\newblock In \emph{Proceedings of the 34th International Conference on Machine
  Learning}, pp.\  2430--2439, 2017.

\bibitem[Pascanu et~al.(2013)Pascanu, Mikolov, and
  Bengio]{DBLP:conf/icml/PascanuMB13}
Pascanu, R., Mikolov, T., and Bengio, Y.
\newblock On the difficulty of training recurrent neural networks.
\newblock In \emph{Proceedings of the 30th International Conference on Machine
  Learning}, pp.\  1310--1318, 2013.

\bibitem[Paszke et~al.(2017)Paszke, Gross, Chintala, Chanan, Yang, DeVito, Lin,
  Desmaison, Antiga, and Lerer]{pytorch}
Paszke, A., Gross, S., Chintala, S., Chanan, G., Yang, E., DeVito, Z., Lin, Z.,
  Desmaison, A., Antiga, L., and Lerer, A.
\newblock Automatic differentiation in pytorch.
\newblock In \emph{NIPS Autodiff Workshop}, 2017.

\bibitem[Real et~al.(2019)Real, Aggarwal, Huang, and
  Le]{DBLP:conf/aaai/RealAHL19}
Real, E., Aggarwal, A., Huang, Y., and Le, Q.~V.
\newblock Regularized evolution for image classifier architecture search.
\newblock In \emph{Proceedings of the 33rd AAAI Conference on Artificial
  Intelligence}, pp.\  4780--4789, 2019.

\bibitem[Seide et~al.(2014)Seide, Fu, Droppo, Li, and
  Yu]{DBLP:conf/interspeech/SeideFDLY14}
Seide, F., Fu, H., Droppo, J., Li, G., and Yu, D.
\newblock 1-bit stochastic gradient descent and its application to
  data-parallel distributed training of speech dnns.
\newblock In \emph{Proceedings of the 15th Conference of the International
  Speech Communication Association}, pp.\  1058--1062, 2014.

\bibitem[Simonyan \& Zisserman(2014)Simonyan and Zisserman]{vgg}
Simonyan, K. and Zisserman, A.
\newblock Very deep convolutional networks for large-scale image recognition.
\newblock \emph{CoRR}, abs/1409.1556, 2014.
\newblock URL \url{http://arxiv.org/abs/1409.1556}.

\bibitem[Strom(2015)]{DBLP:conf/interspeech/Strom15}
Strom, N.
\newblock Scalable distributed {DNN} training using commodity {GPU} cloud
  computing.
\newblock In \emph{Proceedings of the 16th Conference of the International
  Speech Communication Association}, pp.\  1488--1492, 2015.

\bibitem[Sun et~al.(2017)Sun, Zhang, Yu, Zhang, Bhuiyan, and
  Li]{DBLP:conf/ispan/SunZYZBL17}
Sun, C., Zhang, Y., Yu, W., Zhang, R., Bhuiyan, M. Z.~A., and Li, J.
\newblock {DPS:} {A} dsm-based parameter server for machine learning.
\newblock In \emph{Proceedings of the 14th International Symposium on Pervasive
  Systems, Algorithms and Networks {\&} 11th International Conference on
  Frontier of Computer Science and Technology {\&} 3rd International Symposium
  of Creative Computing}, pp.\  20--27, 2017.

\bibitem[Sutskever et~al.(2014)Sutskever, Vinyals, and Le]{seq2seq}
Sutskever, I., Vinyals, O., and Le, Q.~V.
\newblock Sequence to sequence learning with neural networks.
\newblock In \emph{Proceedings of 28th Conference on Neural Information
  Processing Systems}, pp.\  3104--3112, 2014.

\bibitem[Szegedy et~al.(2017)Szegedy, Ioffe, Vanhoucke, and Alemi]{inceptionv4}
Szegedy, C., Ioffe, S., Vanhoucke, V., and Alemi, A.~A.
\newblock Inception-v4, inception-resnet and the impact of residual connections
  on learning.
\newblock In \emph{Proceedings of the 31st {AAAI} Conference on Artificial
  Intelligence}, pp.\  4278--4284, 2017.

\bibitem[Vaswani et~al.(2017)Vaswani, Shazeer, Parmar, Uszkoreit, Jones, Gomez,
  Kaiser, and Polosukhin]{transformer}
Vaswani, A., Shazeer, N., Parmar, N., Uszkoreit, J., Jones, L., Gomez, A.~N.,
  Kaiser, L., and Polosukhin, I.
\newblock Attention is all you need.
\newblock In \emph{Proceedings of the 31st Conference on Neural Information
  Processing Systems}, pp.\  6000--6010, 2017.

\bibitem[Wu et~al.(2017)Wu, Zhu, and E]{DBLP:journals/corr/WuZE17}
Wu, L., Zhu, Z., and E, W.
\newblock Towards understanding generalization of deep learning: Perspective of
  loss landscapes.
\newblock \emph{CoRR}, abs/1706.10239, 2017.
\newblock URL \url{http://arxiv.org/abs/1706.10239}.

\bibitem[Zhang et~al.(2016)Zhang, Gupta, Lian, and
  Liu]{DBLP:conf/ijcai/ZhangGL016}
Zhang, W., Gupta, S., Lian, X., and Liu, J.
\newblock Staleness-aware async-sgd for distributed deep learning.
\newblock In \emph{Proceedings of the 25th International Joint Conference on
  Artificial Intelligence}, pp.\  2350--2356, 2016.

\bibitem[Zhou et~al.(2016)Zhou, Ni, Zhou, Wen, Wu, and
  Zou]{DBLP:journals/corr/ZhouNZWWZ16}
Zhou, S., Ni, Z., Zhou, X., Wen, H., Wu, Y., and Zou, Y.
\newblock Dorefa-net: Training low bitwidth convolutional neural networks with
  low bitwidth gradients.
\newblock \emph{CoRR}, abs/1606.06160, 2016.
\newblock URL \url{http://arxiv.org/abs/1606.06160}.

\end{thebibliography}
\bibliographystyle{sysml2019}

\appendix
%


\end{document}